\documentclass[prodmode,acmtecs]{acmsmalltr} 

\usepackage[ruled]{algorithm2e}

\SetAlFnt{\small}
\SetAlCapFnt{\small}
\SetAlCapNameFnt{\small}
\SetAlCapHSkip{0pt}
\IncMargin{-\parindent}
\usepackage{url}
\usepackage{booktabs}
\usepackage{graphicx}
\usepackage{rotating}
\usepackage{mathrsfs}
\usepackage{color}

\newcommand{\bubing}{BUbiNG\xspace}
\newcommand{\rst}[1]{\ensuremath{{\mathbin\upharpoonright}%
\raise-.5ex\hbox{$#1$}}}

\acmVolume{9}
\acmNumber{4}
\acmArticle{39}
\acmYear{2015}
\acmMonth{1}

\begin{document}
\bibliographystyle{ACM-Reference-Format-Journals}
\markboth{Boldi et al.}{BUbiNG: Massive Crawling for the Masses}

\title{BUbiNG: Massive Crawling for the Masses}
\author{PAOLO BOLDI
\affil{Dipartimento di Informatica, Universit\`a degli Studi di Milano, Italy}
ANDREA MARINO
\affil{Dipartimento di Informatica, Universit\`a degli Studi di Milano, Italy}
MASSIMO SANTINI
\affil{Dipartimento di Informatica, Universit\`a degli Studi di Milano, Italy}
SEBASTIANO VIGNA
\affil{Dipartimento di Informatica, Universit\`a degli Studi di Milano, Italy}}


\begin{abstract}
Although web crawlers have been around for twenty years by now, there is
virtually no freely available, open-source crawling software that guarantees
high throughput, overcomes the limits of single-machine systems and at the same
time scales linearly with the amount of resources available.
This paper aims at filling this gap,
through the description of \bubing, our next-generation web crawler built upon the authors'
experience with UbiCrawler~\cite{BCSU3} and on the last ten years of research
on the topic. \bubing is an open-source Java fully distributed crawler; a single \bubing agent, using sizeable hardware, can crawl
several thousands pages per second respecting strict politeness
constraints, both host- and IP-based.
Unlike existing open-source distributed crawlers that rely on batch techniques
(like MapReduce), \bubing job distribution is based on modern high-speed
protocols so to achieve very high throughput.
\end{abstract}




\acmformat{Paolo Boldi, Andrea Marino, Massimo Santini, and Sebastiano Vigna, 2015. BUbiNG: Massive Crawling for the Masses}

\begin{bottomstuff}
The authors were supported by the EU-FET grant NADINE (GA 288956).

Author's addresses: Via Comelico 39/41, Milano.
\end{bottomstuff}

\maketitle

\section{Introduction}
\label{sec:intro}

A \emph{web crawler} (sometimes also known as a \emph{(ro)bot} or \emph{spider}) is
a tool that downloads systematically a large number of web pages starting
from a seed.
Web crawlers are, of course, used by search engines, but also by companies
selling ``Search--Engine Optimization'' services, by archiving projects such as the Internet
Archive, by surveillance systems (e.g., that scan the web looking for
cases of plagiarism), and by entities performing statistical studies of the
structure and the content of the web, just to name a few.

The basic inner working of a crawler is surprisingly simple from a theoretical viewpoint:
it is a form of traversal (for example, a breadth-first visit). Starting from a
given \emph{seed} of URLs, the set of associated pages is downloaded, their content
is parsed, and the resulting links
are used iteratively to collect new pages.

Albeit in principle a crawler just performs a visit of the web, there are a number of
factors that make the visit of a crawler inherently different from a textbook
algorithm. The first and most important difference is that the size of the graph
to be explored is unknown and huge; in fact, infinite. The second difference is
that visiting a node (i.e., downloading a page) is a complex process that has
intrinsic limits due to network speed, latency, and \emph{politeness}---the requirement of not overloading
servers during the download. Not to mention the countless problems (errors in
DNS resolutions, protocol or network errors, presence of traps) that the crawler
may find on its way.

In this paper we describe the design and implementation of \bubing, our new web
crawler built upon our experience with UbiCrawler~\cite{BCSU3} and on the last
ten years of research on the topic.\footnote{A preliminary poster appeared
in~\cite{BMSVBMCM}.}
\bubing\ aims at filling an important gap in the range of available crawlers. In
particular:
\begin{itemize}
  \item It is a pure-Java, open-source crawler released under the Gnu GPLv3.
  \item It is fully distributed: multiple agents perform the crawl concurrently and
  handle the necessary coordination without the need of any central control;
  given enough bandwidth, the crawling speed grows linearly with the number of agents.
  \item Its design acknowledges that CPUs and OS kernels have become extremely efficient
in handling a large number of threads (in particular, threads that are mainly
I/O-bound) and that large amounts of RAM are by now easily available at a
moderate cost.
More in detail, we assume that the memory used by an agent must be
\emph{constant} in the number of discovered URLs, but that it can \emph{scale linearly} in the
number of discovered hosts. This assumption
simplifies the overall design and makes several data structures more efficient.
\item It is very fast: on a 64-core, 64\,GB workstation it can download hundreds of million of pages at
more than 10\,000 pages per second respecting politeness both by host and by IP, analyzing, compressing and storing
more than 160\,MB/s of data.
\item It is extremely configurable: beyond choosing the sizes of the
various data structures and the communication parameters involved,
implementations can be specified by reflection in a configuration file
and the whole dataflow followed by a discovered URL can be
controlled by arbitrary user-defined filters, that can further be combined with standard Boolean-algebra operators.
\item It fully respects the robot exclusion protocol, a \textit{de facto} standard that well-behaved crawlers are expected to obey.
\item It guarantees that hostwise the visit is an exact breadth-first visit (albeit the global policy
can be customized), thus collecting pages in a more predictible and principled manner.
\item It guarantees that politeness constraints are satisfied both at the host and the IP level, that is,
that two data requests to the same host (name) or IP are separated by at least a
specified amount of time. The two intervals can be set independently, and, in principle, customized per host or IP.
\end{itemize}

When designing a crawler, one should always ponder over the specific usage the
crawler is intended for. This decision influences many of the design details
that need to be taken. Our main goal is to provide a crawler that can be used
out-of-the-box as an archival crawler, but that can be easily modified to
accomplish other tasks. Being an archival crawler, it does not perform any
refresh of the visited pages, and moreover it tries to perform a visit that is
as close to breadth-first as possible (more about this below). Both behaviors
can in fact be modified easily in case of need, but this discussion (on the
possible ways to customize \bubing) is out of the scope of this paper.

We plan to use \bubing to provide new data sets for the research
community. Datasets crawled by UbiCrawler have been used in
hundreds of scientific publications, but \bubing makes it possible
to gather data orders of magnitude larger.

\section{Motivation}

There are four main reasons why we decided to design \bubing as we described above.

\smallskip\noindent\textbf{Principled sampling.} Analyzing the properties of the web graph has
proven to be an elusive goal. A recent large-scale study~\cite{MVLGSW} has shown, once again,
that many alleged properties of the web are actually due to crawling and parsing artifacts instead. By
creating an open-source crawler that enforces a breadth-first visit strategy, altered by
politeness constraints only, we aim at creating web snapshots providing more reproducible results.
While breadth-first visits have their own artifacts (e.g., they can induce an apparent indegree power law
\emph{even on regular graphs}~\cite{ACKBTS}), they are a principled approach that has been widely
studied and adopted. A more detailed analysis, like spam detection, topic selection, and so
on, can be performed offline. A focused crawling activity can actually be \emph{detrimental} to the study
of the web, which should be sampled ``as it is''.

\smallskip\noindent\textbf{Coherent time frame.} Developing a crawler with speed as a main goal
might seem restrictive. Nonetheless, for the purpose of studying the web,
speed is essential, as gathering large snapshots over a long period of time might introduce biases that
would be very difficult to detect and undo.

\smallskip\noindent\textbf{Pushing hardware to the limit.} \bubing is designed to exploit
hardware to its limits, by carefully removing bottlenecks and contention usually present in
highly parallel distributed crawlers. As a consequence, it makes performing large-scale
crawling possible even with limited hardware resources.

\smallskip\noindent\textbf{Consistent crawling and analysis.} \bubing comes
along with a series of tools that make it possible to analyze
the harvested data in a distributed fashion, also exploiting multicore parallelism. In
particular, the construction of the web graph associated with a crawl uses the \emph{same parser as the
crawler}. In the past, a major problem in the analysis of web crawls turned out to be the
inconsistency between the parsing as performed at crawl time and the parsing
as performed at graph-construction time, which introduced artifacts such as spurious components
(see the comments in~\cite{MVLGSW}). By providing a complete framework that uses 
the same code both online and offline we hope to increase the reliability and
reproducibility of the analysis of web snapshots.

\section{Related Works}
Web crawlers have been developed since the very birth of the web. The first-generation 
crawlers date back to the early 90s: World Wide Web Worm~\cite{McBryan1994},
RBSE spider~\cite{Eichmann1994}, MOMspider~\cite{Fielding1994},
WebCrawler~\cite{Pinkerton1994}. One of the main contributions of these works
has been that of pointing out some of the main algorithmic and design issues of
crawlers. In the meanwhile, several commercial search engines, having their own
crawler (e.g., AltaVista), were born.
In the second half of the 90s, the fast growth of the web called for the need of
large-scale crawlers, like the Module crawler~\cite{Burner1997} of the Internet Archive
(a non-profit corporation aiming to
keep large archival-quality historical records of the world-wide web) and the first generation of the
Google crawler~\cite{BrPALHWSE}.
This generation of spiders was able to download efficiently tens of millions of
pages.
At the beginning of 2000, the scalability, the extensibility, and the
distribution of the crawlers become a key design point: this was the case of the
Java crawler Mercator~\cite{NaHHPWC} (the distributed version
of~\cite{HeyN1999}), Polybot~\cite{Shk2002}, IBM WebFountain~\cite{Edwards2001},
and UbiCrawler~\cite{BCSU3}. These crawlers were able to produce snapshots of
the web of hundreds of millions of pages.

Recently, a new generation of crawlers was designed, aiming to download
billions of pages, like~\cite{LeeLWL2009}. Nonetheless, none of them is
freely available and open source: \bubing is the first open-source crawler
designed to be fast, scalable and runnable on commodity hardware.

For more details about previous works or about the main issues in the design of
crawlers, we refer the reader to~\cite{OlstonN10,Mirtaheri2013}.

\subsection{Open-source crawlers}
\label{sec:opensource}

Although web crawlers have been around for twenty years by now (since the spring
of 1993, according to~\cite{OlstonN10}), the area of freely available ones, let
alone open-source, is still quite narrow. With the few exceptions that will be
discussed below, most stable projects we are aware of (GNU \texttt{wget} or mngoGoSearch,
to cite a few) do not (and are not designed to) scale to download more than few
thousands or tens of thousands pages. They can be useful to build an intranet
search engine, but not for web-scale experiments.

Heritrix~\cite{HeritrixWeb,heritrix} is one of the
few examples of an open-source search engine designed to download large datasets: it was developed starting from 2003 by Internet
Archive~\cite{InternetArchive} and it has been since actively
developed. Heritrix (available under the Apache license), although it is of course multi-threaded,
is a single-machine crawler, which is one of the main hindrances to
its scalability. The default crawl order is breadth-first, as suggested by the archival goals behind its design. On the other hand, it provides
a powerful checkpointing mechanism and a flexible way of filtering
and processing URLs after and before fetching.
It is worth noting that the Internet Archive proposed, implemented (in Heritrix) and
fostered a standard format for archiving web content, called WARC, that is now
an ISO standard~\cite{IsoWarc} and that \bubing is also adopting for storing the downloaded
pages.

Nutch~\cite{Khare2004} is one of the best known
existing open-source web crawlers; in fact, the goal of Nutch itself is much
broader in scope, because it aims at offering a full-fledged search engine under
all respects: besides crawling, Nutch implements features such as
(hyper)text-indexing, link analysis, query resolution, result ranking and
summarization. It is natively distributed (using Apache Hadoop as
task-distribution backbone) and quite configurable; it also adopts breadth-first
as basic visit mechanism, but can be optionally configured to go depth-first or
even largest-score first, where scores are computed using some scoring strategy
which is itself configurable.
Scalability and speed are the main design goals of Nutch; for example, Nutch was
used to collect TREC ClueWeb09 dataset\footnote{The new ClueWeb12 dataset,
that is going to be released soon, was collected using Heritrix, instead: five
instances of Heritrix, running on five Dell PowerEdge R410, were run for three
months, collecting $1.2$ billions of pages. The average speed was of about
$38.6$ pages per second per machine.}, the largest web dataset publicly
available as of today consisting of $1\,040\,809\,705$ pages, that were downloaded at the speed of $755.31$ pages/s~\cite{ClueWeb09}; 
to do this they used a Hadoop cluster of 100 machines~\cite{CalSlides}, so their real throughput was of about
$7.55$ pages/s \emph{per machine}. This poor performance is not unexpected: using Hadoop to
distribute the crawling jobs is easy, but not efficient, because it constrains
the crawler to work in a batch\footnote{In theory, Hadoop may perform the prioritization, de-duplication and
distribution tasks \emph{while} the crawler itself is running, but this choice would make the design very
complex and we do not know of any implementation that chose to follow this approach.} fashion. It should not be surprising that using a
modern job-distribution framework like \bubing does increases the throughput by
orders of magnitude.

\section{Architecture overview}

\bubing stands on a few architectural choices which in some cases contrast the
common folklore wisdom. We took our decisions after carefully
comparing and benchmarking several options and gathering the hands-on experience
of similar projects.
\begin{itemize}
  \item The fetching logic of \bubing is built around thousands of identical \emph{fetching threads} performing
   only synchronous (blocking) I/O. Experience with recent Linux kernels and increase in the number of
  cores per machine shows that this approach consistently outperforms
  asynchronous I/O. This strategy simplifies significantly the code complexity,
  and makes it trivial to implement features like HTTP/1.1 ``keepalive'' multiple-resource downloads.
  \item \emph{Lock-free}~\cite{MiSSFPNBBCQA} data structures are used to ``sandwich'' fetching threads, so that they
  never have to access lock-based data structures. This approach is particularly
  useful to avoid direct access to synchronized data structures with logarithmic
  modification time, such as priority queues, as contention between fetching threads can become very significant.
  \item URL storage (both in memory and on disk) is entirely performed using byte arrays. While this approach might seen anachronistic, the Java
  \texttt{String} class can easily occupy three times the memory used by a URL
  in byte-array form (both due to additional fields and to 16-bit characters)
  and doubles the number of objects. \bubing aims at exploiting the large
  memory sizes available today, but garbage collection has a linear cost in the number of objects: this factor must be taken into account.
  \item Following UbiCrawler's design~\cite{BCSU3}, \bubing agents are identical and
  autonomous. The assignment of URLs to agents is entirely customizable, but by
  default we use \emph{consistent hashing} as a fault-tolerant, self-configuring assignment function.
\end{itemize}

\begin{figure*}[tp]
	\centering
	\includegraphics[scale=0.4]{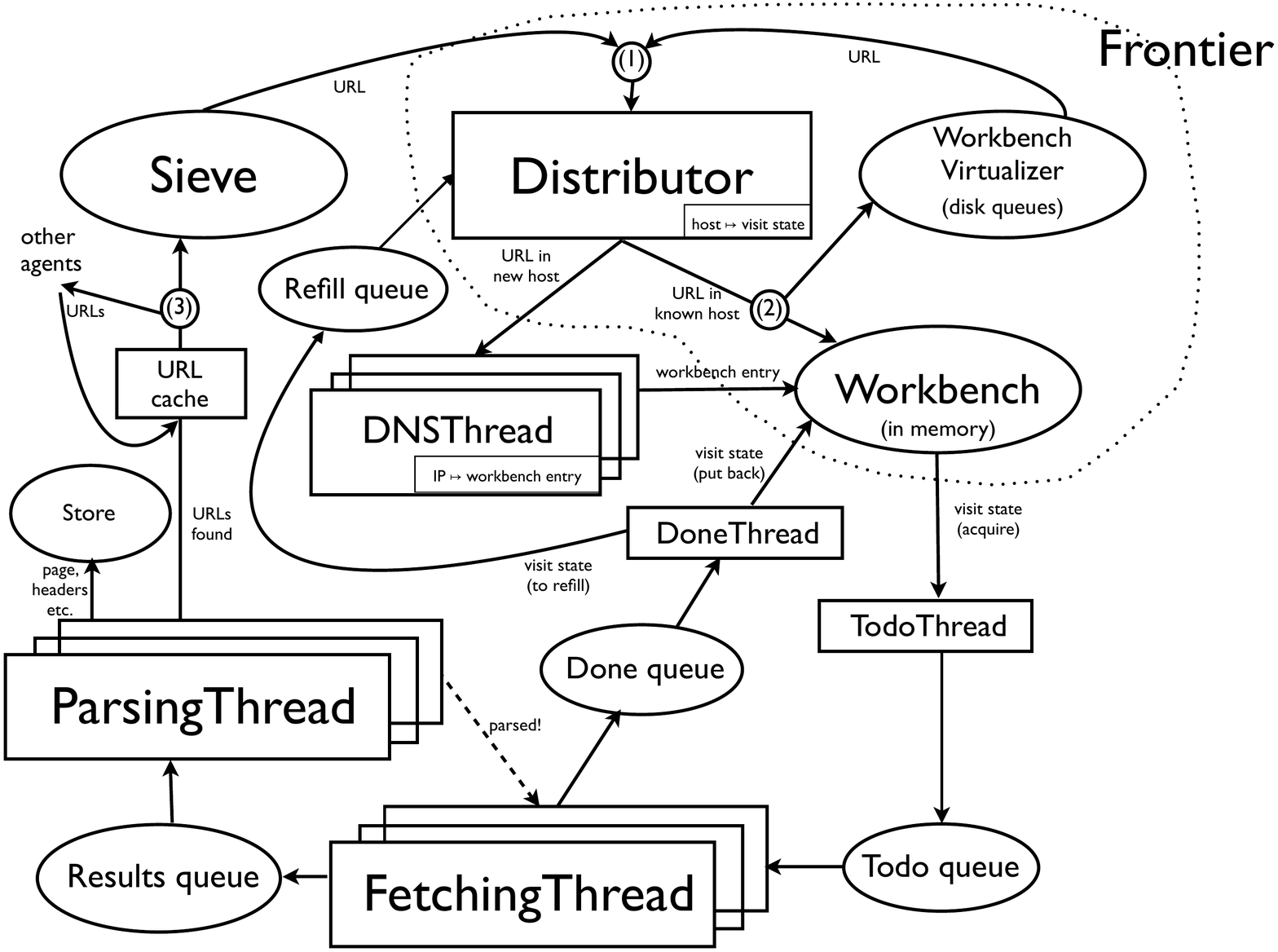}
	\caption{\label{fig:architecture}Overview of the architecture of a \bubing\
	agent. Ovals represent data structures, whereas rectangles represent threads
	(or sets of threads).}
\end{figure*}

In this section, we overview the structure of a \bubing agent: the following sections
detail the behavior of each component. The inner structure and data flow of
an agent is depicted in Figure~\ref{fig:architecture}.

The bulk of the work of an agent is carried out by low-priority \emph{fetching threads}, which
download pages, and \emph{parsing threads}, which parse and extract information
from downloaded pages. Fetching threads are usually thousands, and spend most of their
time waiting for network data, whereas one usually allocates as many
parsing threads as the number of available cores, because their activity is
mostly CPU bound.

Fetching threads are connected to parsing threads using a lock-free \emph{result} list in which
fetching threads enqueue buffers of fetched data, and wait for a parsing thread to analyze them.
Parsing threads poll the result list using an exponential backoff scheme,
perform actions such as parsing and link extraction, and signal back to the fetching thread that
that the buffer can be filled again.

As parsing threads discover new URLs, they enqueue them to a \emph{sieve}
that keeps track of which URLs have been already discovered. A sieve is a data structure similar to a queue with memory:
each enqueued element will be dequeued at some later time, with the guarantee that
an element that is enqueued multiple times will be dequeued just once.
URLs are added to the sieve as they are
discovered by parsing.

In fact, every time a URL is discovered it is checked first against a
high-performance approximate LRU cache containing 128-bit fingerprints: more
than 90\% of the URLs discovered are discarded at this stage. The cache avoids
that frequently found URLs put the sieve under stress, and it has also
another important goal: it avoids that frequently found URLs assigned to another
agent are retransmitted several times.

URLs that come out of the sieve are ready to be visited, and they are taken care
of (stored, organized and managed) by the \emph{frontier}, which is actually
itself decomposed into several modules.

The most important data structure of the frontier is the \emph{workbench}, an
in-memory data structure that keeps track of the visit state of each
host currently being visited and that can check in constant time whether some
host can be accessed for download without violating the politeness constraints. Note that to
attain the goal of several thousands downloaded pages per second without violating politeness constraints
it is necessary to keep track of the visit state of hundreds of thousands of hosts.

When a host is ready for download, its visit state is extracted from the workbench and moved
to a lock-free \emph{todo queue} by a suitable thread. Fetching threads poll
the todo queue with an exponential backoff, fetch resources from the retrieved
visit state\footnote{Possibly multiple resources on a single TCP connection using the ``keepalive'' feature of HTTP 1.1.}
and then put it back onto the workbench. Note that we expect that once
a large crawl has started, the todo queue will never be empty, so fetching threads
will never have to wait. Most of the design challenges of the frontier
components are actually geared towards avoiding that fetching threads ever wait on an empty
todo queue.

The main active component of the frontier is the
\emph{distributor}: it is a high-priority thread that processes URLs coming out
of the sieve (and that must therefore be crawled). Assuming for a moment that
memory is unbounded, the only task of the distributor is that of iteratively dequeueing a URL from
the sieve, checking whether it belongs to a host for which a visit state already
exists, and then either creating a new visit state or enqueuing the URL to an
existing one. If a new visit state is necessary, it is passed to a set of \emph{DNS threads} that
perform DNS resolution and then move the visit state onto the workbench.

Since, however, breadth-first visit queues grow exponentially, and the
workbench can use only a fixed amount of in-core memory, it is necessary to
\emph{virtualize} a part of the workbench, that is, writing on disk part of the URLs
coming out of the sieve.
To decide whether to keep a visit state entirely in the workbench or
to virtualize it, and also to decide when and how URLs should be moved from the
virtualizer to the workbench, the distributor uses a policy that is
described later.

Finally, every agent stores resources in its \emph{store} (that may
possibly reside on a distributed or remote file system).
The native \bubing store is a compressed file in the Web ARChive (WARC)
format (the standard proposed and made popular by Heritrix). This standard
specifies how to combine several digital resources with other information into
an aggregate archive file. In \bubing compression happens in a heavily
parallelized way, with parsing threads compressing independently pages and using
concurrent primitives to pass compressed data to a flushing thread.

\subsection{The sieve}
\label{sec:sieve}

A \emph{sieve} is a queue with memory: it provides enqueue and dequeue primitives,
similarly to a standard queue; each element enqueued to a sieve will be
eventually dequeued later.
However, a sieve guarantees also that if an element is enqueued multiple times,
it will be anyway dequeued just once. Sieves (albeit not called with this name)
have always been recognized as a fundamental basic data structure for a crawler: their main
implementation issue lies in the unbounded, exponential growth of the number of discovered URLs. While it is
easy to write enqueued elements to a disk file, checking that an element is not
returned multiple times requires \textit{ad-hoc} data structures---standard
dictionaries would use too much in-core memory.

The actual sieve implementation used by \bubing can be customized, but the
default one, called \texttt{MercatorSieve}, is similar to the one
suggested in~\cite{HeyN1999} (hence its name).
Each element known to the sieve is stored as a 64-bit hash in a disk file.
Every time a new element is enqueued, its hash is stored in an in-memory array,
and the element is saved in an auxiliary file.
When the array is full, it is
sorted and compared with the set of elements known to the sieve. The auxiliary file
is then scanned, and previously unseen elements are stored for later dequeueing.
All these operations require only sequential access to all files involved, and
the sizing of the array is based on the amount of in-core memory available. 
Note that the output order is guaranteed to be the same of the input order
(i.e., elements are appended in the order of their first appearance).

A generalization of the idea of a sieve, with the additional possibility of
associating values with the elements, is the DRUM (Disk Repository with Update
Management) structure used by IRLBot and described in~\cite{LeeLWL2009}. A DRUM
provides additional operations to retrieve or update the values associated with
the elements.
From an implementation viewpoint, DRUM is a Mercator sieve with multiple arrays,
called \emph{buckets}, in which a careful orchestration of in-memory and on-disk
data makes it possible to sort in one shot sets that are an order of magnitude larger
than what the Mercator sieve would allow using the same quantity of in-core
memory.
However, to do so DRUM must sacrifice breadth-first order: due to the inherent
randomization of the way keys are placed in the buckets, there is no guarantee
that URLs will be crawled in breadth-first order, not even per host.
Finally, the tight analysis in~\cite{LeeLWL2009} about the properties of DRUM is
unavoidably bound to the single-agent approach of IRLBot: for example, the
authors conclude that a URL cache is not useful to reduce the number of
insertions in the DRUM, but the same cache reduces significantly network transmissions.
Based on our experience, once the cache is in place the Mercator sieve becomes much more competitive.

There are several other implementations of the sieve logic currently used. A
quite common choice is to adopt an explicit queue and a \emph{Bloom
filter}~\cite{BloSTTOHCAW} to remember enqueued elements. Albeit popular, this
choice has no theoretical guarantees: while it is possible to decide \textit{a
priori} the maximum number of pages that will ever be crawled, it is very
difficult to bound in advance the number of \emph{discovered} URLs, and this
number is essential in sizing the Bloom filter. If the discovered URLs are significantly
more than expected, several pages are likely to be lost because of false
positives.
A better choice is to use a dictionary of fixed-size \emph{fingerprints} obtained
from URLs using a suitable hash function. The disadvantage is that the structure
would no longer use constant memory.

\subsection{The workbench}
\label{sec:workbench}

The \emph{workbench} is an in-memory data structure that contains the next URLs
to be visited. It is one of the main novel ideas in \bubing's design, and it is
one of the main reasons why we can attain a very high throughput. It is a
significant improvement over IRLBot's two-queues approach~\cite{LeeLWL2009}, as
it can detect in constant time whether a URL is
ready for download without violating politeness limits.

First of all, URLs associated with a specific host\footnote{Every
URL is made~\cite{Lee2005} by a scheme (also popularly called
``protocol''), an authority (a host, optionally a port number, and perhaps some
user information) and a path to the resource, possibly followed by a query
(that is separated from the path by a ``?'').
\bubing's data structures are built around the pair scheme+authority, but in this paper we
will use the more common
word ``host'' to refer to it.} are kept in a structure
called \emph{visit state}, containing a FIFO queue of the next URLs to be
crawled for that host along with a \texttt{next-fetch} field that specifies the
first instant in time when a URL from the queue can be downloaded, according to
the per-host politeness configuration. Note that inside a visit state we
only store a byte-array representation of the path and query of a URL:
this approach significantly reduces object creation,
and provides a simple form of compression by prefix omission.

Visit states are further grouped into \emph{workbench entries} based on their IP address;
every time the first URL for a given host is found, a new visit state is created
and then the IP address is determined (by one of the \emph{DNS threads}): the
new visit state is either put in a new workbench entry (if no known host was as
associated to that IP address yet), or in an existing one.

A workbench entry contains a queue of visit states (associated with the same IP) prioritized by their
\texttt{next-fetch} field, and an IP-specific \texttt{next-fetch}, containing the first
instant in time when the IP address can be accessed again, according to the
per-IP politeness configuration.
The \emph{workbench} is the queue of all workbench entries, prioritized on the
\texttt{next-fetch} field of each entry \emph{maximized} with the \texttt{next-fetch}
field on the top element of its queue of visit states.
In other words, the workbench is a priority queue of priority queues of FIFO queues.

We remark that due to our choice of priorities \emph{there is a host that can be
visited without violating host or IP politeness constraints if and only if the first URL of
the top visit state of the top workbench entry can be visited}. Moreover, if
there is no such host, the delay after which a host will be ready is given by
the priority of the top workbench entry minus the current time.

Therefore, the workbench acts as a \emph{delay queue}: its dequeue operation waits, if necessary,
until a host is ready to be visited. At that point, the top entry $E$ is removed from the workbench and the top
visit state is removed from $E$. Both removals happen in logarithmic time (in the number of visit states).
The visit state and the associated workbench entry act as a \emph{token} that is
virtually passed between \bubing's components to guarantee that no component is
working on the same workbench entry at the same time (in particular, this forces
both kinds of politeness). In practice, as we mentioned in the overview,
dequeueing is performed by a high-priority
thread, the \emph{todo thread}, that constantly dequeues visit states from the workbench and enqueues them
into a lock-free \emph{todo queue}, which is then accessed by fetching threads. This approach,
besides avoiding contention by thousands of threads on a relatively slow
structure, makes the number of visit states that are ready for downloads easily measurable: it is just
the size of the todo queue. The downside is that, in principle, using very skewed per-host or per-IP politeness
delays might cause the order of the todo queue not to reflect the actual priority of the visit
states contained therein.


\subsection{Fetching threads}

A \emph{fetching thread} is a very simple thread that iteratively extracts visit
states from the todo queue.
If the todo queue is empty, a standard exponential backoff procedure is used to
avoid polling the list too frequently, but the design of \bubing aims at keeping
the todo queue nonempty and avoiding backoff altogether.

Once a fetching thread acquires a visit state, it tries to fetch the first
URL of the visit state FIFO queue. If suitably configured, a fetching thread can
also iterate the fetching process on more URLs for a fixed amount of time, so
to exploit the ``keepalive'' feature of HTTP 1.1.

Each fetching thread has an associated \emph{fetch data} instance in which
the downloaded data are buffered. Fetch data include a transparent buffering
method that keeps a fixed amount of data in memory and dumps on disk the remaining part.
By sizing the fixed
amount suitably, most requests can be completed without accessing the disk, but at
the same time rare large requests can be handled without allocating additional memory.

After a resource has been fetched, the fetch data is put in the
\emph{results} queue so that one of the parsing threads can parse it. Once this process
is over, the parsing thread sends a signal back so that the fetching thread is able
to start working on a new URL. Once a fetching thread
has to work on a new visit state, it puts the current visit state in a \emph{done queue},
from which it will be dequeued by a suitable thread
that will then put it back on the workbench together with its associated entry.

Most of the time, a fetching thread
is blocked on I/O, which makes it possible to run thousands of them in parallel.
Indeed, the number of fetching threads determines the amount of parallelization
\bubing can achieve while fetching data from the network, so it should be chosen
as large as possible, compatibly with the amount of bandwidth available and with
the memory used by fetch data.

\subsection{Parsing threads}

A \emph{parsing thread} iteratively extracts from the results queue
the fetch data that have been previously enqueued by a fetching thread.
Then, the content of the HTTP response is analyzed and possibly parsed.
If the response contains a HTML page, the parser will produce
a set of URLs that will be first checked against the URL cache, and then, if not already seen,
either sent to another agent, or enqueued to the same agent's sieve.

During the parsing phase, a parsing thread computes a digest of the response
content. The signature is stored in a Bloom filter~\cite{BloSTTOHCAW} and it is
used to avoid saving several times the same page (or near-duplicate pages).
Finally, the content of the response is saved to the store.

Since two pages are considered (near-)duplicates whether they have the same
signature, the digest computation is responsible for content-based duplicate
detection.
In the case of HTML pages, in order to collapse near-duplicates, some heuristic
is used. In particular, an hash fingerprint is computed on a summarized content,
which is obtained by stripping HTML attributes, and discarding digits and dates
from the response content. This simple heuristic allows for instance to collapse
pages that differs just for visitor counters or calendars. In a post-crawl
phase, there are several more sophisticated approaches that can be applied, like
\emph{shingling} \cite{Broder1997}, \emph{simhash}
\cite{Charikar02}, \emph{fuzzy fingerprinting} \cite{Fetterly03,Chakrabarti2003},
and others, e.g.,~\cite{Manku2007}.

For the sake of description, we will refer to duplicates as the pages which are
(near-)duplicates of some other page previously crawled according to the above
definition, while we will call archetypes the set of pages which are not
duplicates.

\subsection{DNS threads}

DNS threads are used to solve host names of new hosts:
a DNS thread continuously dequeues from the list of newly
discovered visit states and \emph{resolves} its host name, adding it to a
workbench entry (or creating a new one, if the IP address itself is new), and putting
it on the workbench. In our experience, it is essential to run a local recursive
DNS server to avoid the bottleneck caused by an external server.

\subsection{The workbench virtualizer}

The workbench virtualizer maintains on disk a mapping from hosts to FIFO \emph{virtual queues}
of URLs. Conceptually, all URLs that have been extracted from the sieve but have not yet
been fetched are enqueued in the workbench visit state they belong to, in the exact
order in which they came out of the sieve. Since, however, we aim at crawling
with an amount of memory that is \emph{constant} in the number of discovered
URLs, part of the queues must be written on disk. Each virtual queue contains
a fraction of URLs from each visit state, in such a way that the overall
URL order respects, \emph{per host}, the original breadth-first order.

Virtual queues are consumed as the visit proceeds, following the natural per-host breadth-first
order. As fetching threads download URLs, the workbench is partially freed and can be filled
with URLs coming from the virtual queues. This action is performed by the same
thread emptying the done queue (the queue containing the visit states after
fetching): as it puts visit states back on the workbench, it selects visit
states with URLs on disk but no more URLs on the workbench and puts them on a
\emph{refill queue} that will be later read by the distributor.

Initially, we experimented with virtualizers inspired by the BEAST module of
IRLbot~\cite{LeeLWL2009}, although many crucial details of their implementation were missing (e.g., the treatment
of HTTP and connection errors); moreover, due to the
static once-for-all distribution of URLs among a number of physical on-disk
queues, it was impossible to guarantee adherence to a breadth-first visit in the
face of unpredictable network-related faults.

Our second implementation was based on the Berkeley DB, a key/value store that
is also used by Heritrix. While extremely popular, Berkeley DB is a
general-purpose storage system, and in particular in Java it has a very heavy load in terms of
object creation and corresponding garbage collection. While providing in principle services
like URL-level prioritization (which was not one of our design goals),
Berkeley DB was soon detected to be a serious bottleneck in the overall design.

We thus decided to develop an \textit{ad-hoc} virtualizer oriented towards
breadth-first visits. We borrowed from Berkeley DB the idea of writing data in log files
that are periodically collected, but we decided to rely on memory mapping to lessen
the I/O burden.

In our virtualizer, on-disk URL queues are stored in log files that are memory mapped and transparently thought of
as a contiguous memory region.
Each URL stored on disk is prefixed with a pointer to the position of the next URL
for the same host. Whenever we append a new URL, we
modify the pointer of the last stored URL for the same host accordingly. A small amount
of metadata associated with each host (e.g., the head and tail of its queue) is stored in main memory.

As URLs are dequeued to fill the workbench, part of the log files become free. When
the ratio between the used and allocated space goes below a threshold (e.g.,
$50\%$), a garbage-collection process is started. Due to the fact that URLs are
always appended, there is no need to keep track of free space: we just scan the
queues in order of first appearance in the log files and gather them at the
start of the memory-mapped space.
By keeping track (in a priority queue) of the position of
the next URL to be collected in each queue, we can move items directly to their final position,
updating the queue after each move. We stop when enough space has been freed, and delete
the log files that are now entirely unused.

Note that most of the activity of our virtualizer is caused by appends
and garbage collections (reads are a lower-impact activity that is necessarily bound
by the network throughput). Both activities are highly localized (at the end of the
currently used region in the case of appends, and at the current collection point in the
case of garbage collections), which makes a good use of the caching facilities of the operating system.

\subsection{The distributor}
\label{sec:distributor}

The \emph{distributor} is a high-priority thread that orchestrates the movement
of URLs out of the sieve, and loads URLs from virtual queues into the workbench
as necessary.

As the crawl proceeds, URLs get accumulated in visit states at
different speeds, both because hosts have different responsiveness and because
websites have different sizes and branching factors.
Moreover, the workbench has a (configurable) limit size that
cannot be exceeded, since one of the central design goals of \bubing is that the
amount of main memory occupied cannot grow unboundedly in the number of the
discovered URLs, but only in the number of hosts discovered.
Thus, filling the workbench blindly with URLs coming out of the sieve would soon
result in having in the workbench only URLs belonging to a limited number of
hosts.

The \emph{front} of a crawl, at any given time, is the number of visit states that are ready
for download respecting the politeness constraints. The front size determines the overall throughput of the
crawler---because of politeness, the number of distinct hosts currently
being visited is the crucial datum that establishes how fast or slow the crawl
is going to be.

One of the two forces driving the distributor
is, indeed, that \emph{the front should always be large enough so that no fetching thread
has ever to wait}.
To attain this goal, the distributor enlarges dynamically the \emph{required
front size}: each time a fetching thread has to wait, albeit the current front
size is larger than the current required front size, the latter is increased.
After a warm-up phase, the required front size stabilizes to a value that
depends on the kind of hosts visited and on the amount of resources available.
At that point, it is impossible to have a faster crawl given the resources available,
as all fetching threads are continuously downloading data. Increasing the number of
fetching threads, of course, may cause an increase of the required front size.

The second force driving the distributor is the (somewhat informal) requirement that \emph{we
try to be as close to a breadth-first visit as possible}. Note that this force works
in an opposite direction with respect to enlarging the front---URLs that are already in
existing visit states should be in principle visited \emph{before} any URL in the sieve,
but enlarging the front requires dequeueing more URLs from the sieve to find new hosts.

The distributor is also responsible for filling the workbench with URLs coming either
out of the sieve, or out of virtual queues (circle
numbered $(1)$ in Figure~\ref{fig:architecture}). Once again, staying close to a breadth-first
visit requires loading URLs in virtual queues, but keeping the front large might
call for reading URLs from the sieve to discover new hosts.

The distributor privileges refilling the queues of the workbench using URLs from
the virtualizer, because this makes the visit closer to an exact breadth-first. However, if no
refill has to be performed and the front is not large enough,
the distributor will read from the sieve, hoping to find new hosts to make the front larger.

When the distributor reads a URL from the sieve, the URL can either be 
put in the workbench or written in a virtual queue, depending on whether there
are already URLs on disk for the same host, and on the number of URLs per IP address
that should be in the workbench to keep it full, but not overflowing, when the front
is of the required size.

\begin{figure}[tp]
	\centering
	\includegraphics[scale=0.35]{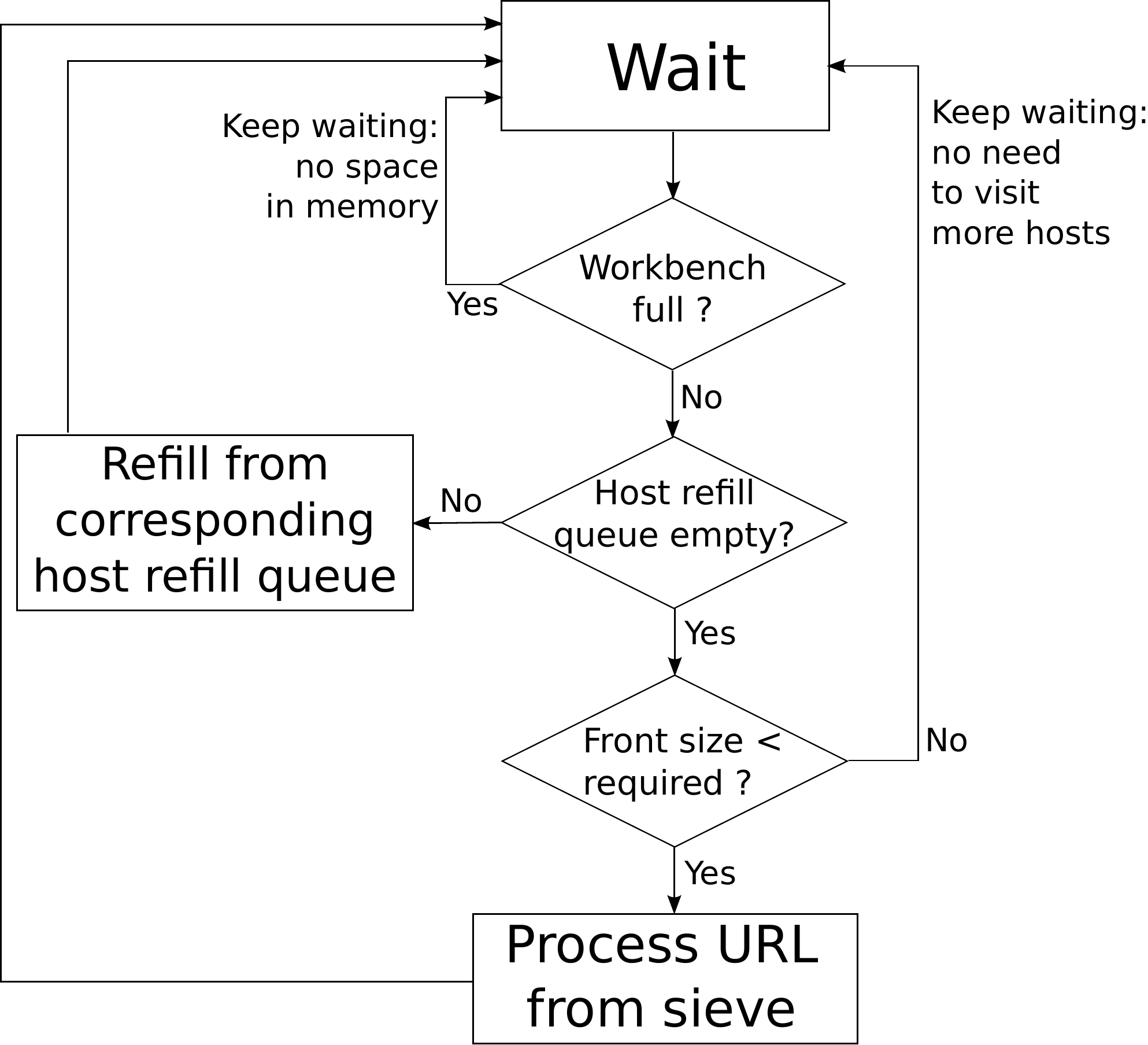}
	\caption{\label{fig:flowchart}How the distributor interacts with the sieve,
	the workbench and the workbench virtualizer.}
\end{figure}


\subsection{Configurability}
\label{sec:conf}

To make \bubing capable of a versatile set of tasks and behaviors, every
crawling phase (fetching, parsing, following the URLs of a page, scheduling new
URLs, storing pages) is controlled by a \emph{filter}, a Boolean predicate that
determines whether a given resource should be accepted or not. Filters can be
configured both at startup and at runtime allowing for a very fine-grained
control.

The type of objects a filter considers is called the \emph{base type} of the
filter. In most cases, the base type is going to be a URL or a fetched page.
More precisely, a \emph{prefetch} filter is one that has a \bubing URL as its
base type (typically: to decide whether a URL should be scheduled for later
visit, or should be fetched); a \emph{postfetch} filter is one that has a
fetched response as base type and decides whether to do something with that
response (typically: whether to parse it, to store it, etc.).

\subsection{Heuristics}
\label{sec:heur}

Some classes of BUbiNG contain the distillation of heuristics we
developed in almost twenty years of work with web crawling. 

One important such class is \texttt{BURL} (a short name for ``BUbiNG URL''),
which is the class responsible for parsing and normalizing URLs found in web
pages. The topic of parsing and normalization is much more involved than one
might expect---very recently, the failure in building a sensible web graph from
the ClueWeb09 collection stemmed in part from the lack of suitable
normalization of the URLs involved. \texttt{BURL} takes care of fine details
such as escaping and de-escaping (when unnecessary) of non-special characters,
case normalization of percent-escape.

\subsection{Distributed crawling}
\label{sec:distribution}

\bubing crawling activity can be distributed by running several agents over
multiple machines. Similarly to UbiCrawler~\cite{BCSU3}, all agents are
identical instances of \bubing, without any explicit leadership:
all data structures described above are part of each agent.

URL assignment to agents is entirely configurable. By default, \bubing uses just
the host to assign a URL to an agent, which avoids that two different agents can
crawl the same host at the same time.
Moreover, since most hyperlinks are local, each agent will be himself
responsible for the large majority of URLs found in a typical HTML
page~\cite{OlstonN10}.
Assignment of hosts to agents is by default performed using \emph{consistent
hashing}~\cite{BCSU3}.

Communication of URLs between agents is handled by the message-passing methods
of the JGroups Java library; in particular, to make communication lightweight
URLs are by default distributed using UDP. More sophisticated
communications between the agents rely on the TCP-based JMX Java standard remote-control
mechanism, which exposes most of the internal configuration parameters and statistics.
Almost all crawler structures are indeed modifiable at runtime.

\begin{table}
	\tbl{Comparison between \bubing and the main
	existing open-source crawlers. Resources are HTML pages for ClueWeb09 and IRLBot, but include
	other data types (e.g., images) for ClueWeb12.  For reference, we also report the throughput of
	IRLbot~\cite{LeeLWL2009}, although the latter is not open source. Note that ClueWeb09 was gathered using
	a heavily customized version of Nutch.}{
	\begin{tabular}{|l|r|r|r|r|r|r|r|}
	\hline
		\multicolumn{1}{|c|}{ } &
		\multicolumn{1}{c|}{ } &
		\multicolumn{1}{c|}{Resources} &
		\multicolumn{2}{c|}{Resources/s} &
		\multicolumn{2}{|c|}{Speed in MB/s} \\
\cline{4-7}
		\multicolumn{1}{|c|}{Crawler} &
		\multicolumn{1}{c|}{Machines} &
		\multicolumn{1}{c|}{(Millions)} &
		\multicolumn{1}{c|}{overall} &
		\multicolumn{1}{c|}{per agent} &
		\multicolumn{1}{c|}{overall}&
		\multicolumn{1}{c|}{per agent} \\
	\hline
	\hline

		Nutch (ClueWeb09) &
		100 (Hadoop) & 
		$1\,200$ & 
		$430$ & 
		$4.3$ & 
		$10$ &
		$0.1$\\ 
	\hline
		%
		Heritrix (ClueWeb12) &
		5 & 
		$2\,300$ & 
		$300$ & 
		$60$ & 
		$19$ &
		$4$ \\  
	Heritrix (\textit{in vitro}) &
		$1$ &
		$115$ &
		$370$ & 
		$370$ & 
		$4.5$ &		
		$4.5$ \\  
	\hline
		IRLBot &
		1 &
		$6\,380$ & 
		$1\,790$ & 
		$1\,790$ &
		$40$ &
		$40$ \\ 
	\hline
		\bubing (iStella) &
		1 &
		500 &
		$3\,700$ &
		$3\,700$ &
		$154$ &
		$154$ \\
		\bubing (\textit{in vitro}) &
		$4$ &
		$1\,000$ &
	 	$40\,600$ &
		$10\,150$ &
		$640$ &		
		$160$ \\
 	\hline
	\end{tabular}
	}
	\label{tab:comp}
\end{table}

\section{Experiments}

Testing a crawler is a delicate, intricate, arduous task: on one hand, every
real-world experiment is obviously influenced by the hardware at one's disposal
(in particular, by the available bandwidth). Moreover, real-world tests are
difficult to repeat many times with different parameters: you will either end up
disturbing the same sites over and over again, or choosing to visit every time a
different portion of the web, with the risk of introducing artifacts in the
evaluation. Given these considerations, we ran two kinds of experiments: one
batch was performed \textit{in vitro} with a HTTP proxy\footnote{The proxy
software is distributed along with the rest of \bubing.} simulating network
connections towards the web and generating fake HTML pages
(with a configurable behavior that includes delays, protocol exceptions etc.),
and another batch of experiments was performed \emph{in vivo}.

\subsection{\textit{In vitro} experiments: \bubing}

To verify the robustness of \bubing\ when varying some basic parameters, such as
the number of fetching threads or the IP delay, we decided to run some \textit{in
vitro} simulations on a group of four machines sporting 64 cores and 64\,GB of
core memory. In all experiments, the number of parsing and DNS threads was
fixed and set respectively to 64 and 10. The size of the workbench was set
to 512MB, while the size of the sieve was set to 256MB. We
always set the host politeness delay equal to the IP politeness delay.
Every \textit{in
vitro} experiment was run for 90 minutes.

\begin{figure}[tp]
\centering
\includegraphics[scale=.65]{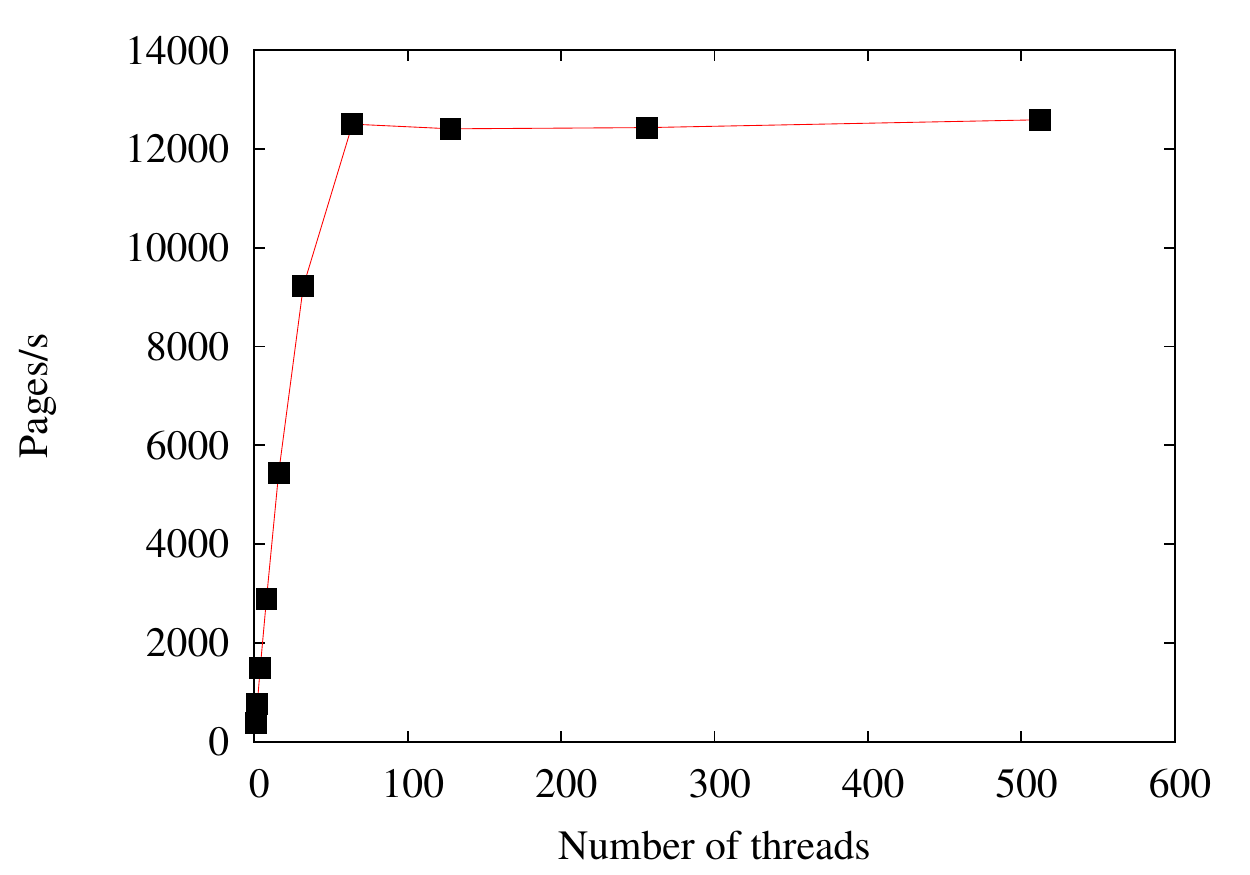}
\caption{\label{fig:threadsvsspeed} The average number of pages per second with respect to the number of threads
using a simulated slow connection. Note the linear increase in speed until the plateau, due to the limited (300) number of
threads of the proxy.}
\end{figure}

\smallskip
\noindent\textbf{Fetching threads.}
The first thing we wanted to test was that
increasing the number of fetching threads yields a better usage of the network, and hence a larger number of
requests per second, until the bandwidth is saturated. The results of this experiment are
shown in Figure~\ref{fig:threadsvsspeed} and have been obtained by having the
proxy simulate a network that saturates quickly, using no politeness delay.
The behavior visible in the plot tells us that the increase in the number of
fetching threads yields a linear increase in speed until the
available (simulated) bandwidth is reached; after that, the number of requests
stabilizes to a plateau. Also this part of the plot tells us something: after
saturating the bandwidth, we do not see any decrease in the throughput,
witnessing the fact that our infrastructure does not cause any hindrance to the
crawl.

\begin{figure}[tp]
\centering
\includegraphics[scale=.65]{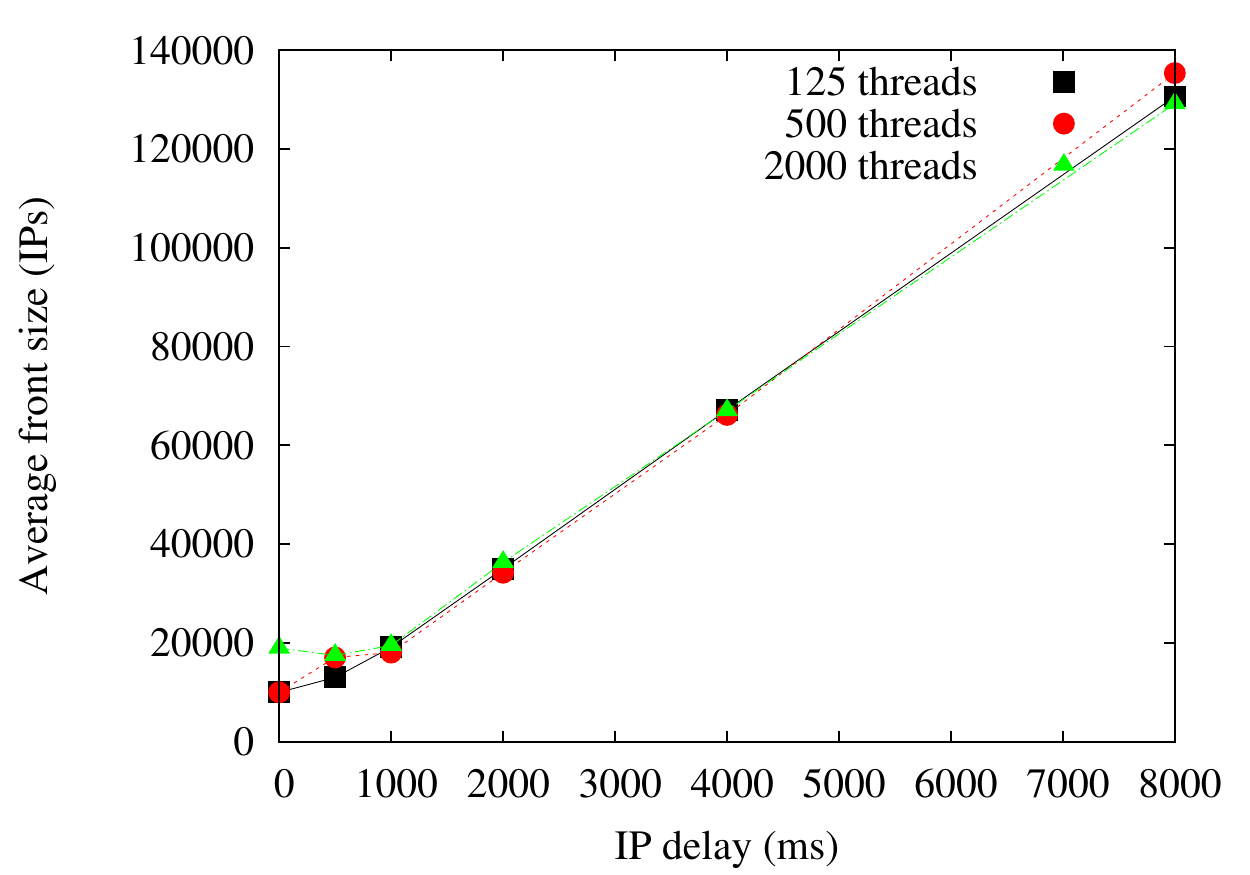}
\includegraphics[scale=.65]{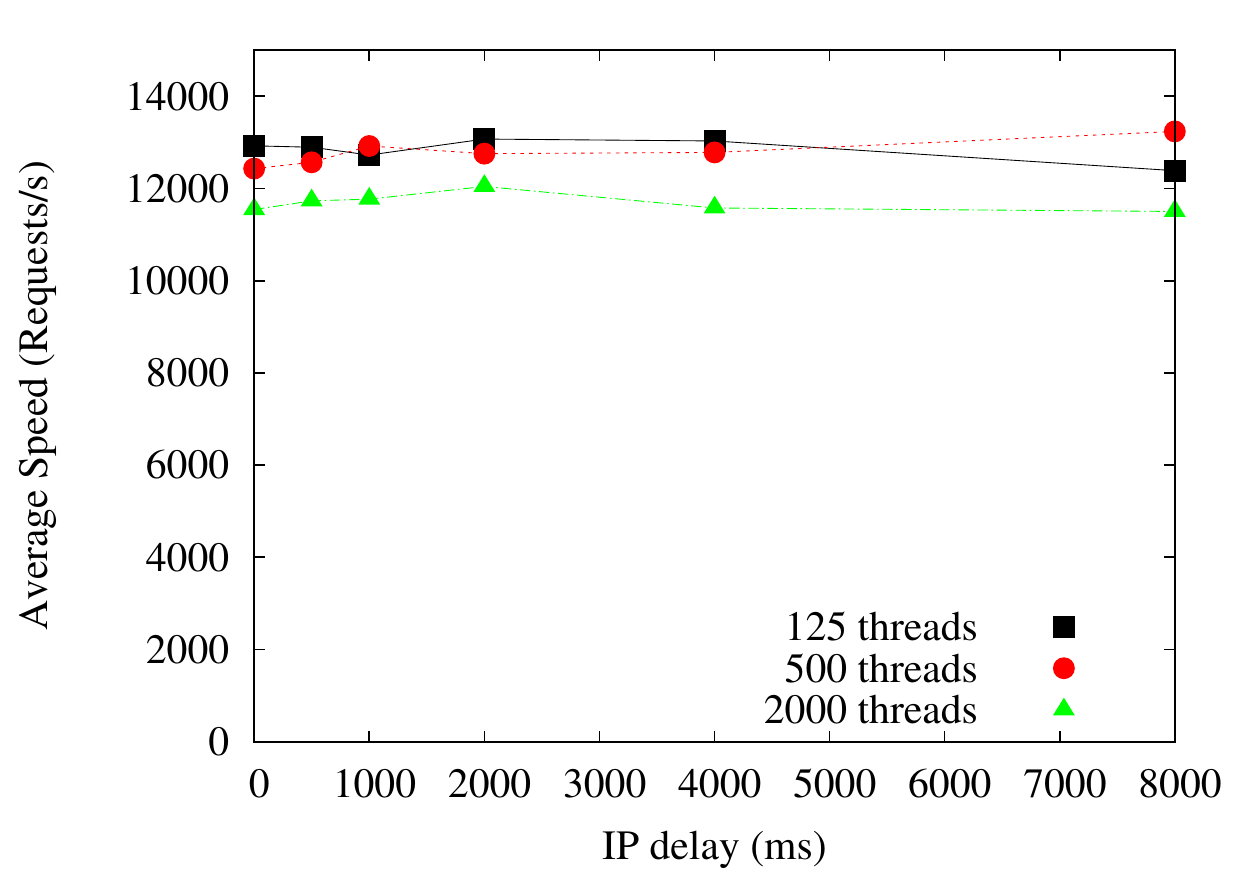}
\includegraphics[scale=.65]{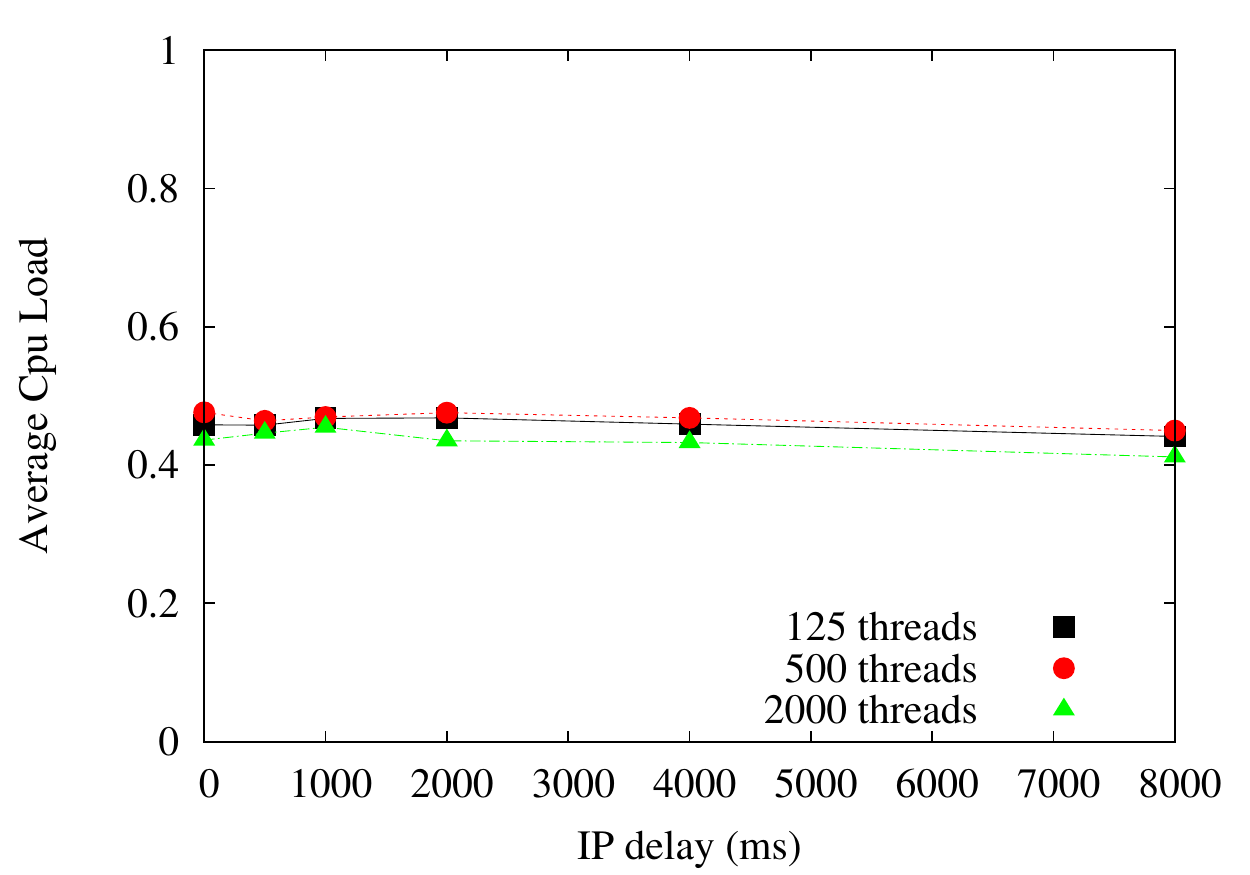}
\caption{\label{fig:politeness} The average size of the front, the average number of requests per second, and the average CPU load with respect to the IP delay (the host delay is set to eight times the IP delay). Note that the front adapts linearly to the growth of the IP delay, and, due to the essentially unlimited bandwidth of the proxy, the number of fetching threads is almost irrelevant.}
\end{figure}

\smallskip
\noindent\textbf{Politeness.}
The experiment described so far uses a small number of fetching
threads, because the purpose was to show what happens before saturation.
Now we show what happens under a heavy load.
Our second \textit{in vitro} experiment keeps the number of fetching threads fixed but
increases the amount of politeness, as determined by the IP delay. We plot \bubing's throughput as the IP delay (hence the host delay) increases
in Figure~\ref{fig:politeness} (top): to maintain the same
throughput, the front size (i.e., the number of hosts being visited in parallel)
must increase, as expected. Moreover, this is independent on the number of
threads (of course, until the network bandwidth is saturated). In the same
figure we show that the average throughput is independent from the politeness (and almost independent
from the number of fetching threads), and the same is true of the CPU load.
Even if this fact might seem surprising,
this is the natural consequence of two observations:
first, even with a small number of fetching threads, \bubing\ always tries to
fill the bandwidth and to maximize its usage of computational resources;
second, even varying the IP and host delay, \bubing\ modifies the number of hosts
under visit to tune the interleaving between their processing.

\smallskip \noindent\textbf{Raw speed.} We wanted to test the raw speed of a
cluster of \bubing agents. We thus ran four agents using 1\,000 fetching
threads, until we gathered one
billion pages, averaging 40\,600 pages per second on the whole cluster.
We also ran the same test on a single machine, obtaining essentially the same
per-machine speed, showing that \bubing scales linearly with the number of
agents in the cluster.

\smallskip \noindent\textbf{Testing for bottlenecks: no I/O.} Finally, we wanted
to test whether our lock-free architecture was actually able to sustain a very
high parallelism. To do so, we ran a no-I/O test on a 40-core workstation. The purpose of the test was to
stress the computation and contention bottlenecks in absence of any interference from
I/O: thus, input from the network was generated internally using the same logic
of our proxy, and while data was fully processed (e.g., compressed)
no actual storage was performed. After 100 million pages, the average speed
was $16\,000$ pages/s (peak $22\,500$) up to 6\,000 threads. We detected the
first small decrease in speed ($15\,300$ pages/s, peak $20\,500$) at 8\,000
threads, which we believe is physiological due to increased context switch and Java garbage collection.

\subsection{\textit{In vitro} experiments: Heritrix}

To provide a comparison of \bubing with another crawler in a completely equivalent setting,
we ran a raw-speed test using Heritrix 3.2.0 on the same hardware as in the
\bubing raw-speed experiment, always using a proxy with the same setup. We configured Heritrix to use the same amount of memory,
$20$\% of which was reserved for the Berkeley DB cache. We used 1\,000 threads,
locked the politeness interval to 10 seconds regardless of the download time (by default, Heritrix uses an adaptive scheme),
and enabled content-based duplicate detection.\footnote{We thank Gordon Mohr,
one of the authors of Heritrix, for suggesting us how to configure it for a large workstation.}
The results obtained will be presented and discussed in
Section~\ref{sec:comparison}.


\subsection{\textit{In vivo} experiments}

We performed a number of experiments \textit{in vivo} at different sites. The
main problem we had to face is that a single \bubing agent on sizable hardware
can saturate a 1\,Gb/s geographic link, so, in fact, we were not initially able
to perform any test in which the network was not capping the crawler. Finally,
iStella, an Italian commercial search engine provided us with a 48-core, 512\,GB
RAM with a 2\,Gb/s link. The results confirm the knowledge we have gathered with
our \textit{in vitro} experiment: in the iStella experiment we were able to
keep a steady download speed of $1.2$\,Gb/s using a single \bubing agent crawling
the \texttt{.it} domain. The overall CPU load was about $85$\%.


\subsection{Comparison}
\label{sec:comparison}

When comparing crawlers, many measures are possible, and depending on the task at hand, different measures might be
suitable. For instance, crawling all types of data (CSS, images, etc.) usually yields a significantly higher throughput than
crawling just HTML, since HTML pages are often rendered dynamically, sometimes
causing a significant delay, whereas most other types are served statically. The crawling policy has also a huge influence on the throughput: prioritizing by indegree (as IRLBot does~\cite{LeeLWL2009}) or alternative
importance measure shifts most of the crawl on sites hosted on powerful servers with large-bandwidth connection. Ideally,
crawlers should be compared on a crawl with given number of pages in
breadth-first fashion from a fixed seed, but some crawlers are not available to the public, which makes this goal unattainable.

In Table~\ref{tab:comp} we gather some evidence of the excellent performance of \bubing.
Part of the data is from the literature, and part has been generated during our experiments.

First of all, we report performance data for Nutch and Heritrix from the recent
crawls made for the ClueWeb project (ClueWeb09 and ClueWeb12). The figures
are those available in~\cite{CalSlides} along with those found
in~\cite{ClueWeb09} and~\url{http://boston.lti.cs.cmu.edu/crawler/crawlerstats.html}:
notice that the data we have about those collections are sometimes slightly
contradictory (we report the best figures). The comparison with the ClueWeb09 crawl is somewhat
unfair (the hardware used for that dataset was ``retired search-engine
hardware''), whereas the comparison with ClueWeb12 is more unbiased, as the hardware used was more recent.
We report the throughput declared by IRLBot~\cite{LeeLWL2009},
too, albeit the latter is not open source and the downloaded data is not
publicly available.

Then, we report experimental \textit{in vitro} data about Heritrix and \bubing
obtained, as explained in the previous section, using the same hardware, a
similar setup, and a HTTP proxy generating web pages.\footnote{Note that, with the purpose of stress testing the crawler internals, our
HTTP proxy generates fairly short pages. This feature explains the wildly different
ratio between MB/s and resources/s when looking at \textit{in vitro} and
\textit{in vivo} experiments.} This figures are the ones that can be compared
more appropriately. Finally, we report the data of the iStella experiment.

The results of the comparison show quite clearly that the speed of \bubing is
several times that of IRLBot and one to two orders of magnitude larger than
that of Heritrix or Nutch. 

All in all, our experiments show that \bubing's adaptive design provides a very
high throughput, in particular when a strong politeness is desired: indeed, from
our comparison, the highest throughput.
The fact that the throughput can be scaled linearly just by adding agents makes
it by far the fastest crawling system publicly available.


\section{Three datasets}

As a stimulating glimpse into the capabilities of BUbiNG to collect interesting datasets,
we describe the main features of three snapshots collected with different criteria. All snapshots
contain about one billion unique pages (the actual crawls are significantly larger, due to duplicates).
\begin{itemize}
  \item \texttt{uk-2014}: a snapshot of the \texttt{.uk} domain, taken with a limit of 10\,000 pages per host starting from the BBC website.
  \item \texttt{eu-2015}: a ``deep'' snapshot of the national domains of the
  European Union, taken with a limit of 10\,000\,000 pages per host starting from \texttt{europa.eu}.
  \item \texttt{gsh-2015}: a general ``shallow'' worldwide snapshot, taken with a limit of 100 pages per host, always
  starting from \texttt{europa.eu}.
\end{itemize}
The \texttt{uk-2014} snapshot follows the tradition of our laboratory of taking
snapshots of the \texttt{.uk} domain for linguistic uniformity, and to obtain a regional snapshot. The second and third snapshot aims at exploring the 
difference in the degree distribution and in website centrality in two very different kinds of data-gathering activities.
In the first case, the limit on the pages per host is so large that, in fact, it
was never reached; it is a quite faithful ``snowball sampling'' due to the breadth-first nature of BUbiNG's visits. In the second case,
we aim at maximizing the number of collected hosts by downloading very few pages
per host. One of the questions we are trying to answer using the latter two snapshots is: how much is the
indegree distribution dependent on the cardinality of sites (root pages have an indegree usually at least as large as the site size), and how
much is it dependent on inter-site connections?

The main data, and some useful statistics about the three datasets, are shown in
Table~\ref{tbl:basic}. Among these, we have the average number of links per page
(average outdegree) and the average number of links per page whose destination is
on a different host (average external outdegree). Moreover, concerning the graph
induced by the pages of our crawls, we also report the average distance, the
harmonic diameter (e.g., the harmonic mean of all the distances), and the
percentage of reachable pairs of pages in this graph (e.g., pairs of nodes $(x,y)$ for which there exists
a directed path from $x$ to $y$).

\subsection{Degreee distribution}

\begin{figure}[tp]
\centering
\begin{tabular}{ccccc}
\includegraphics[scale=0.35]{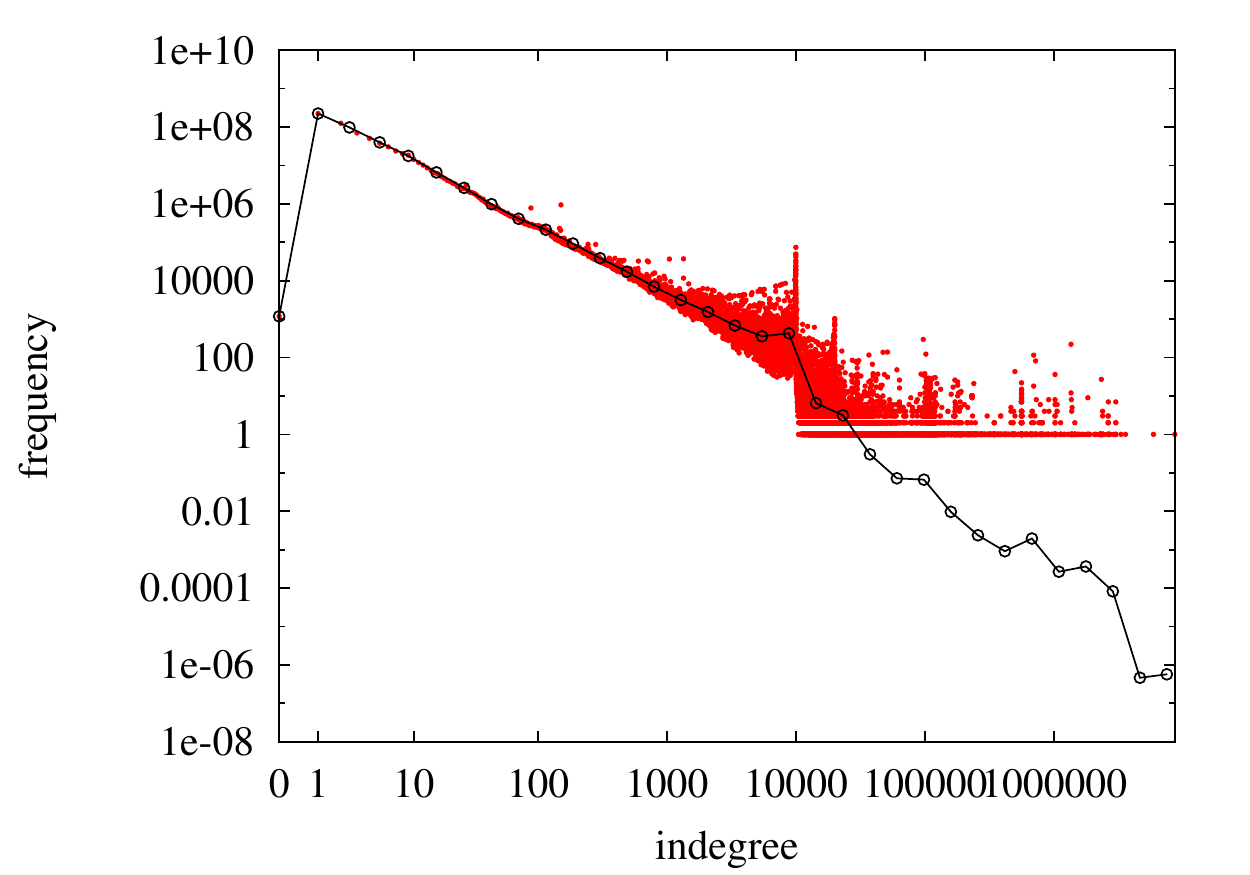} 
& &
\includegraphics[scale=0.35]{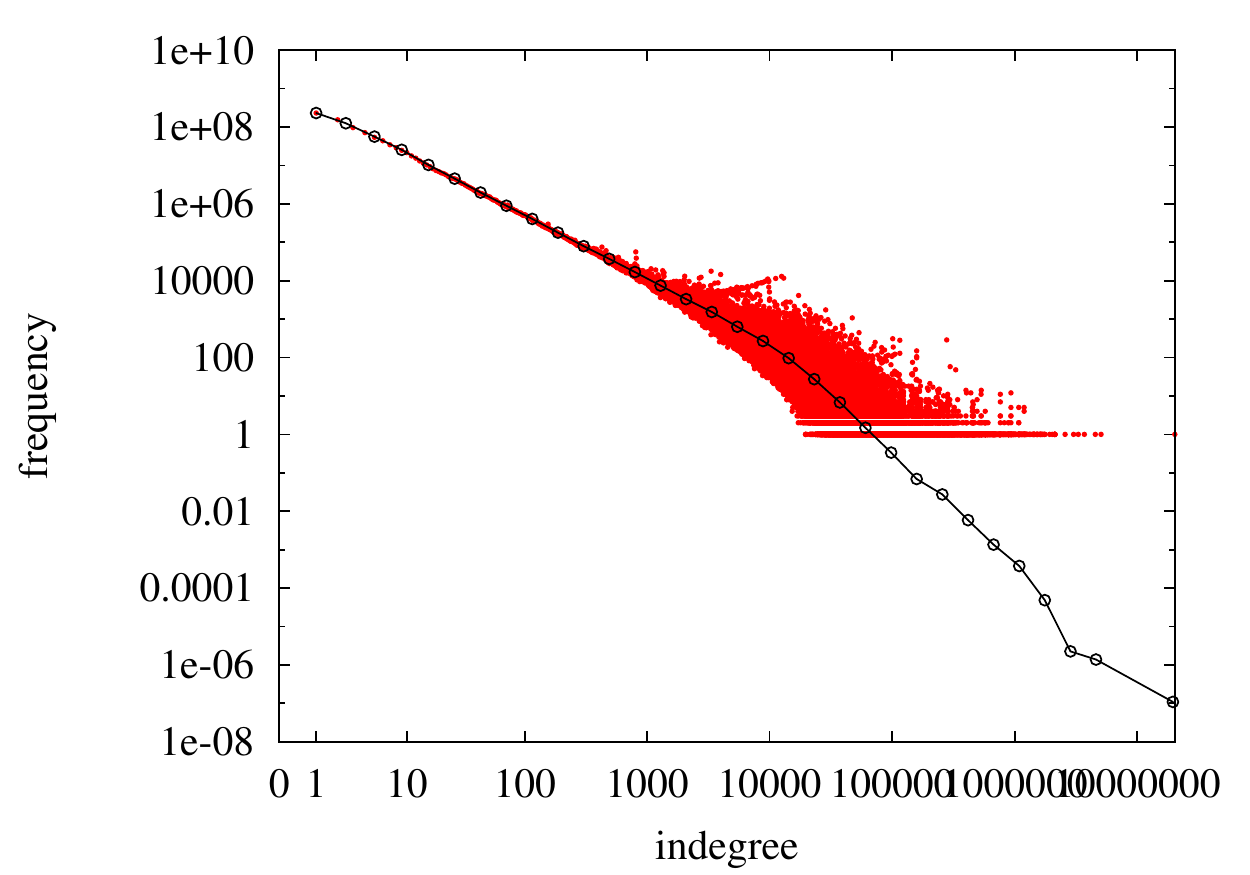}
& &
\includegraphics[scale=0.35]{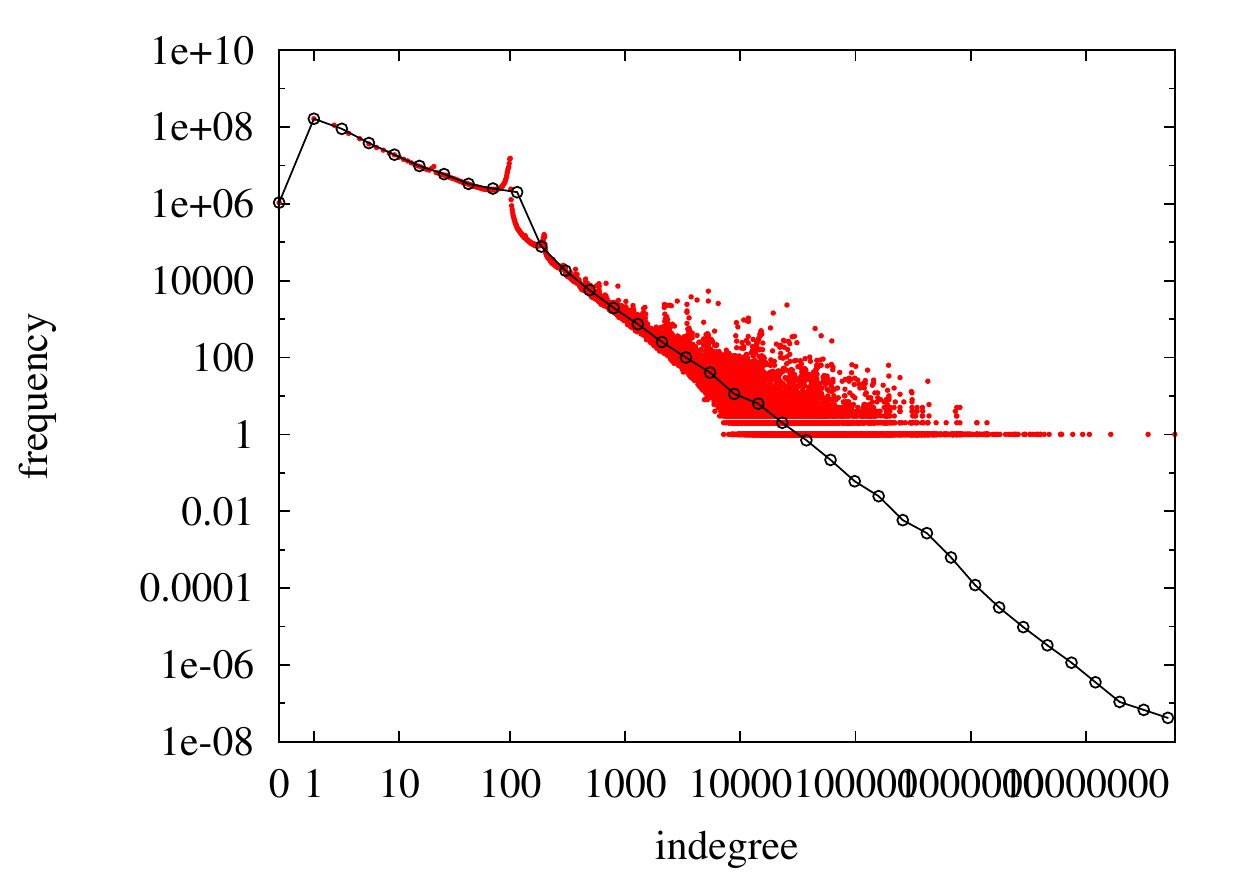}
\end{tabular}
\caption{\label{fig:indeg}Indegree plots for \texttt{uk-2014}, \texttt{eu-2015} and \texttt{gsh-2015} (degree/frequency plots with Fibonacci binning).}
\end{figure}

\begin{figure}[tp]
\centering
\begin{tabular}{ccccc}
\includegraphics[scale=0.35]{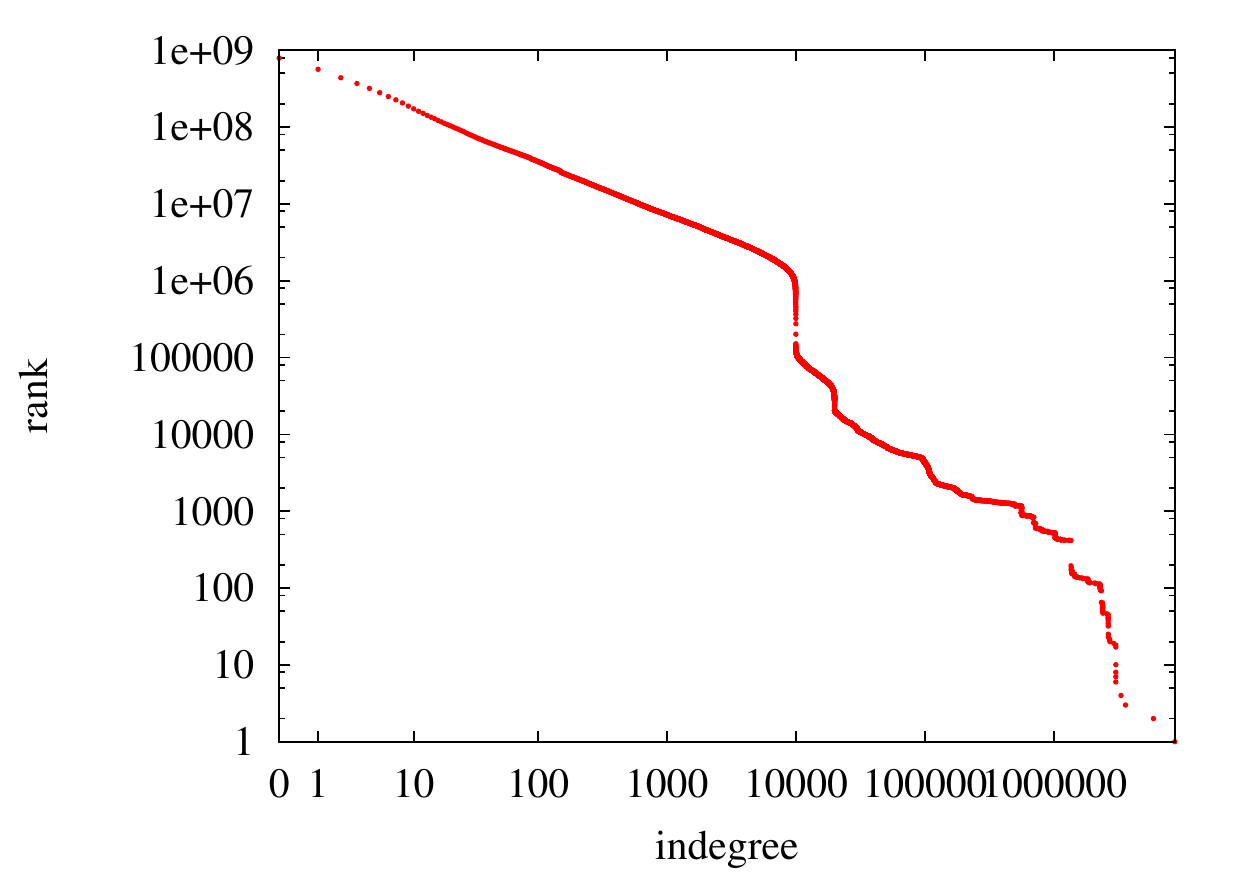} 
& &
\includegraphics[scale=0.35]{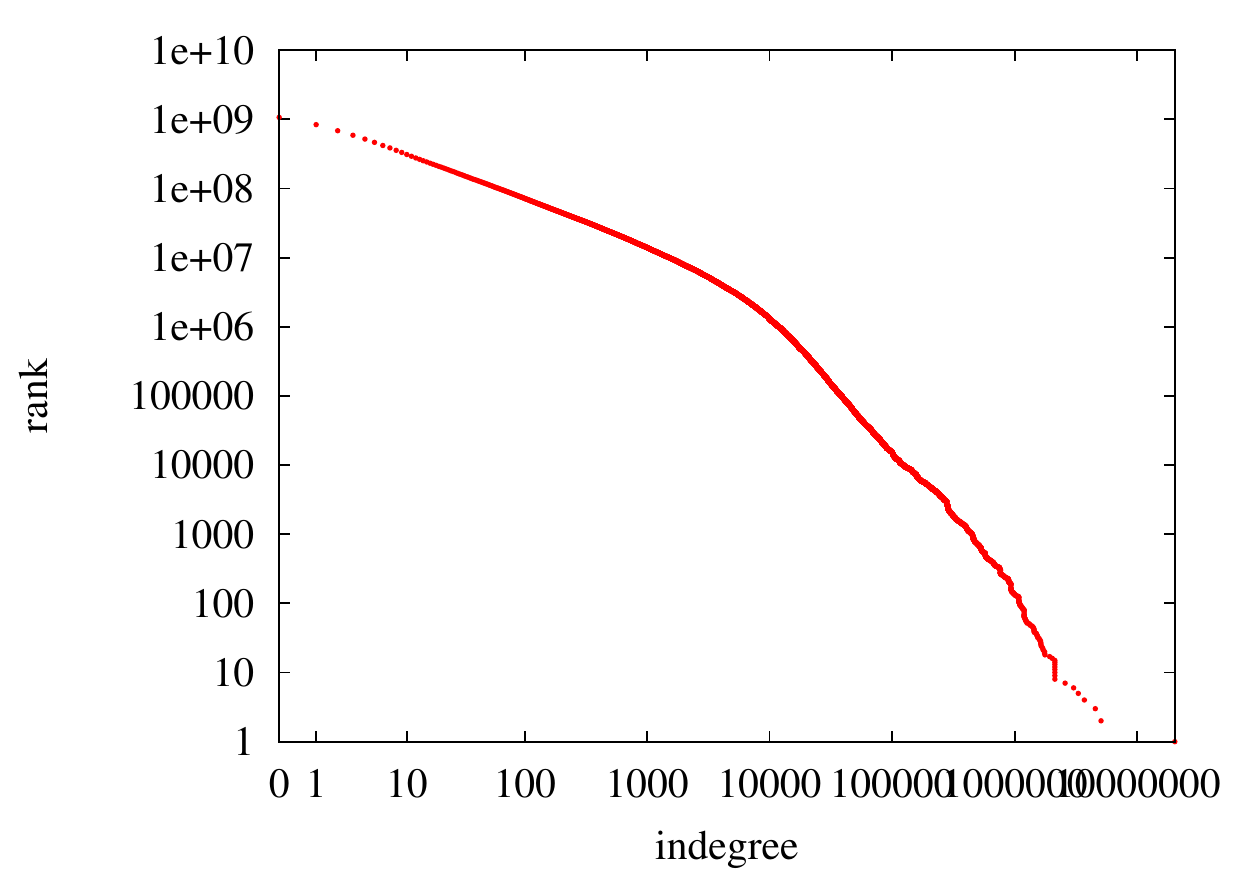}
& &
\includegraphics[scale=0.35]{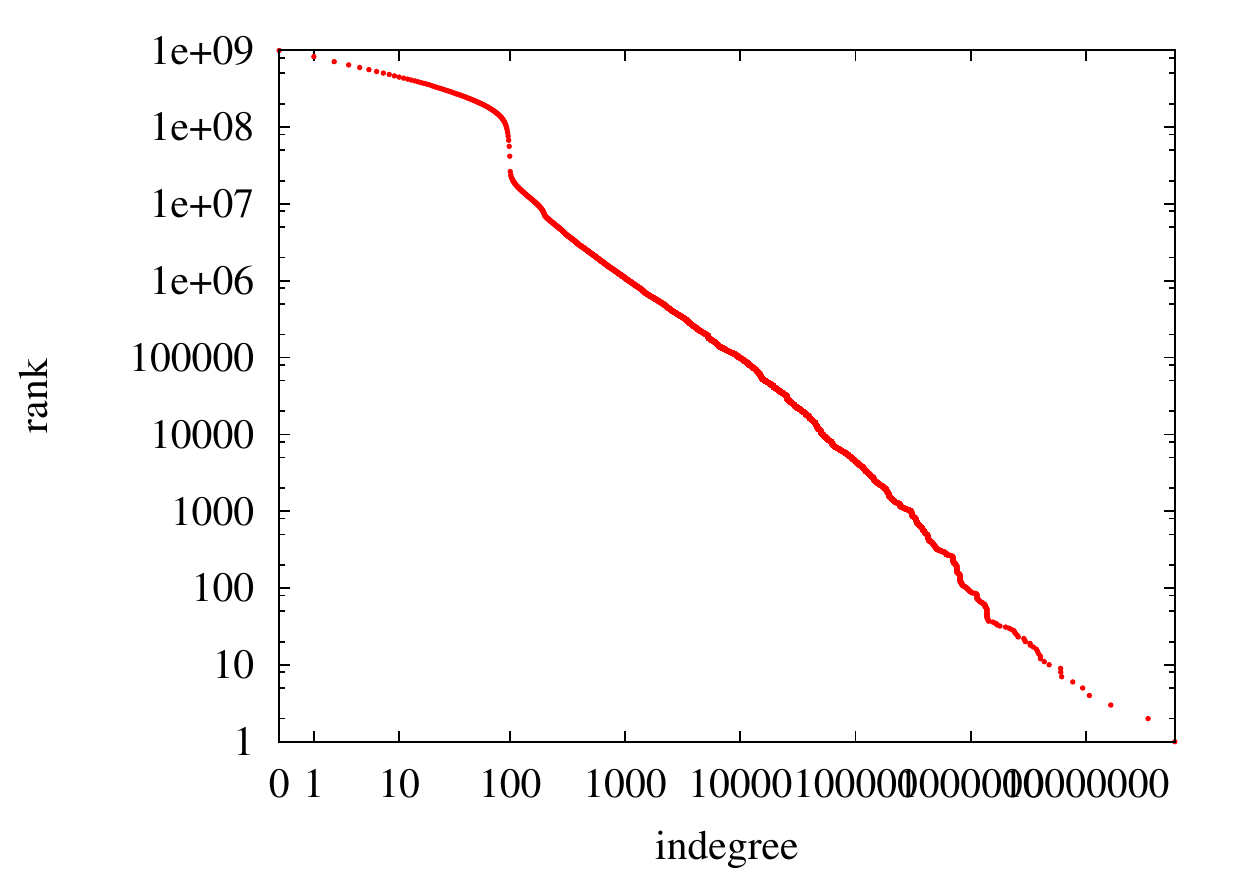}
\end{tabular}
\caption{\label{fig:indeg-sr}Indegree plots for \texttt{uk-2014}, \texttt{eu-2015} and \texttt{gsh-2015} (cumulative degree/rank plots).}
\end{figure}

\begin{figure}[tp]
{\centering
\begin{tabular}{ccccc}
\includegraphics[scale=0.35]{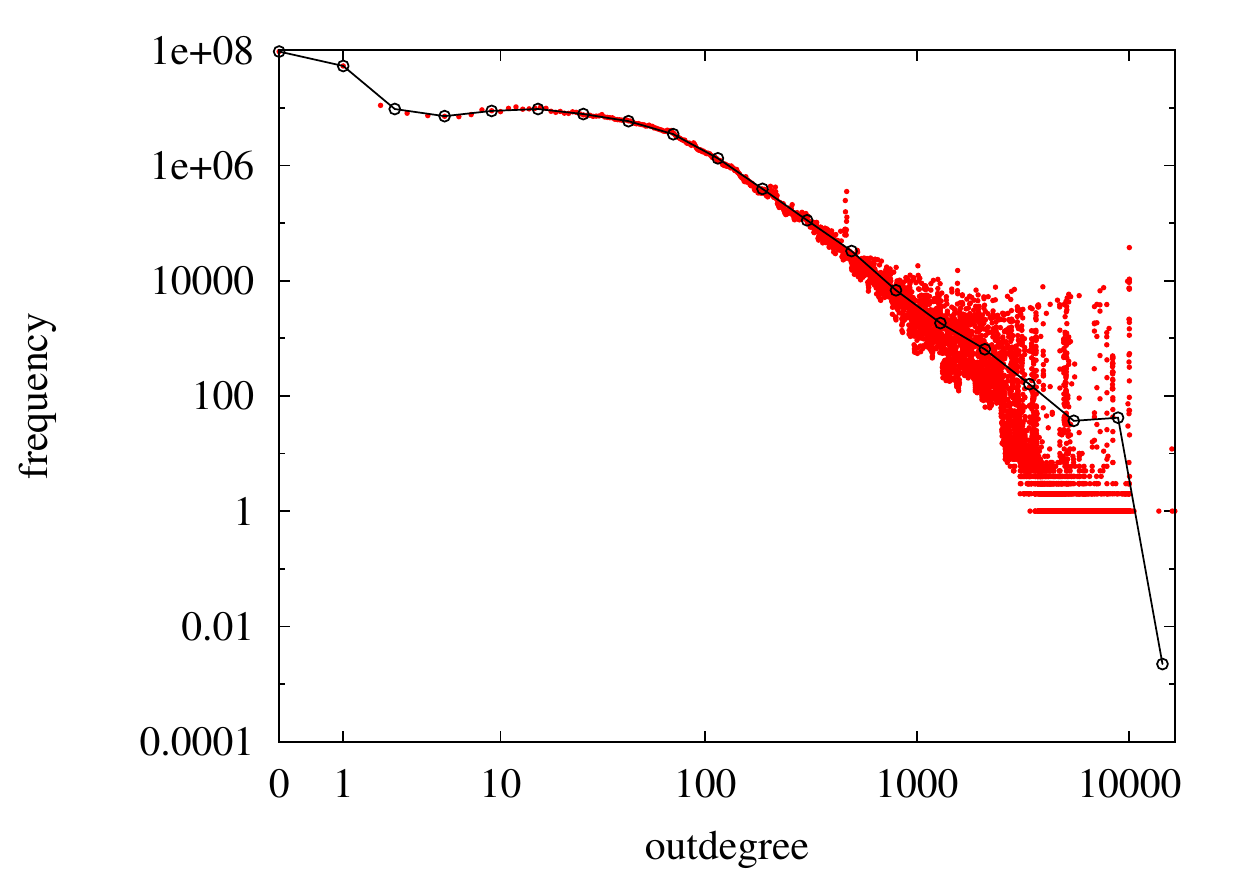} 
& &
\includegraphics[scale=0.35]{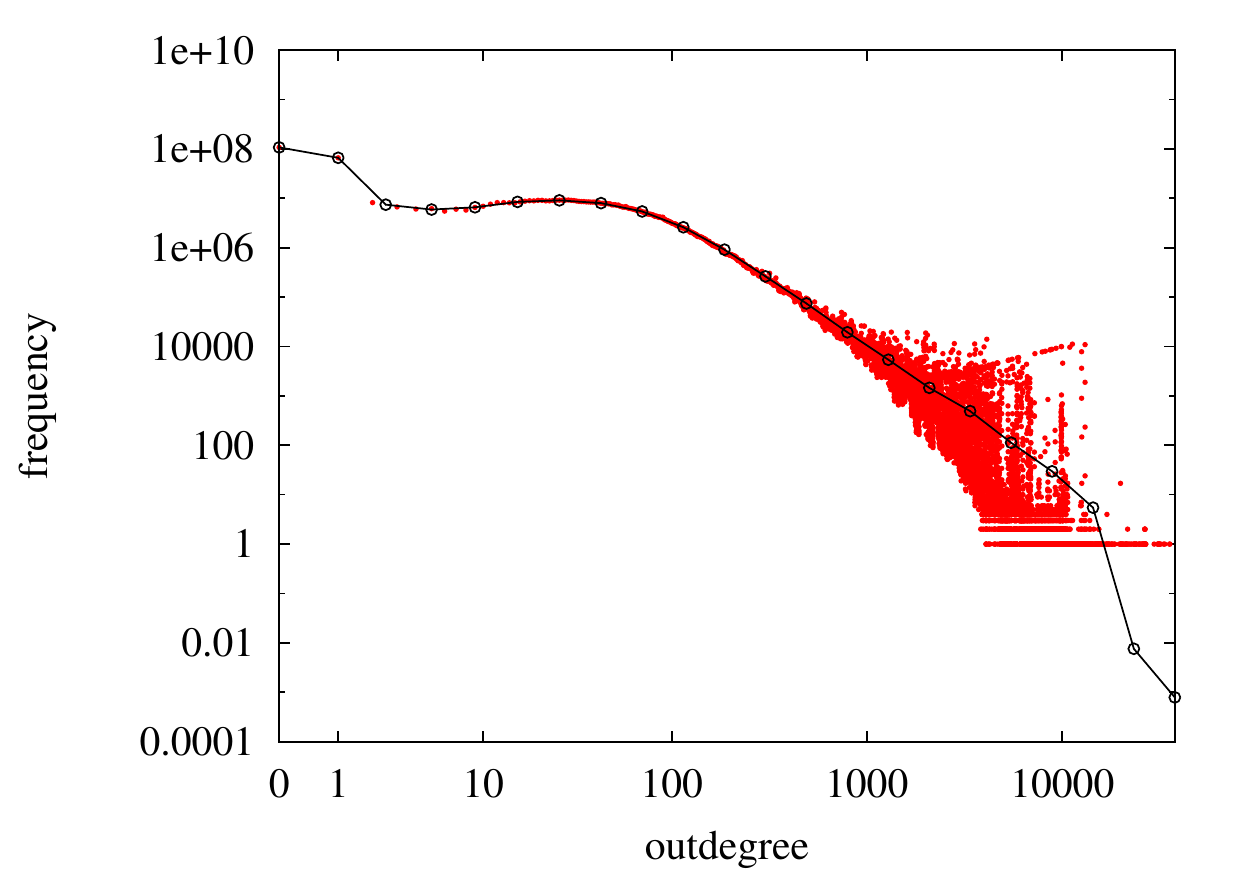}
& &
\includegraphics[scale=0.35]{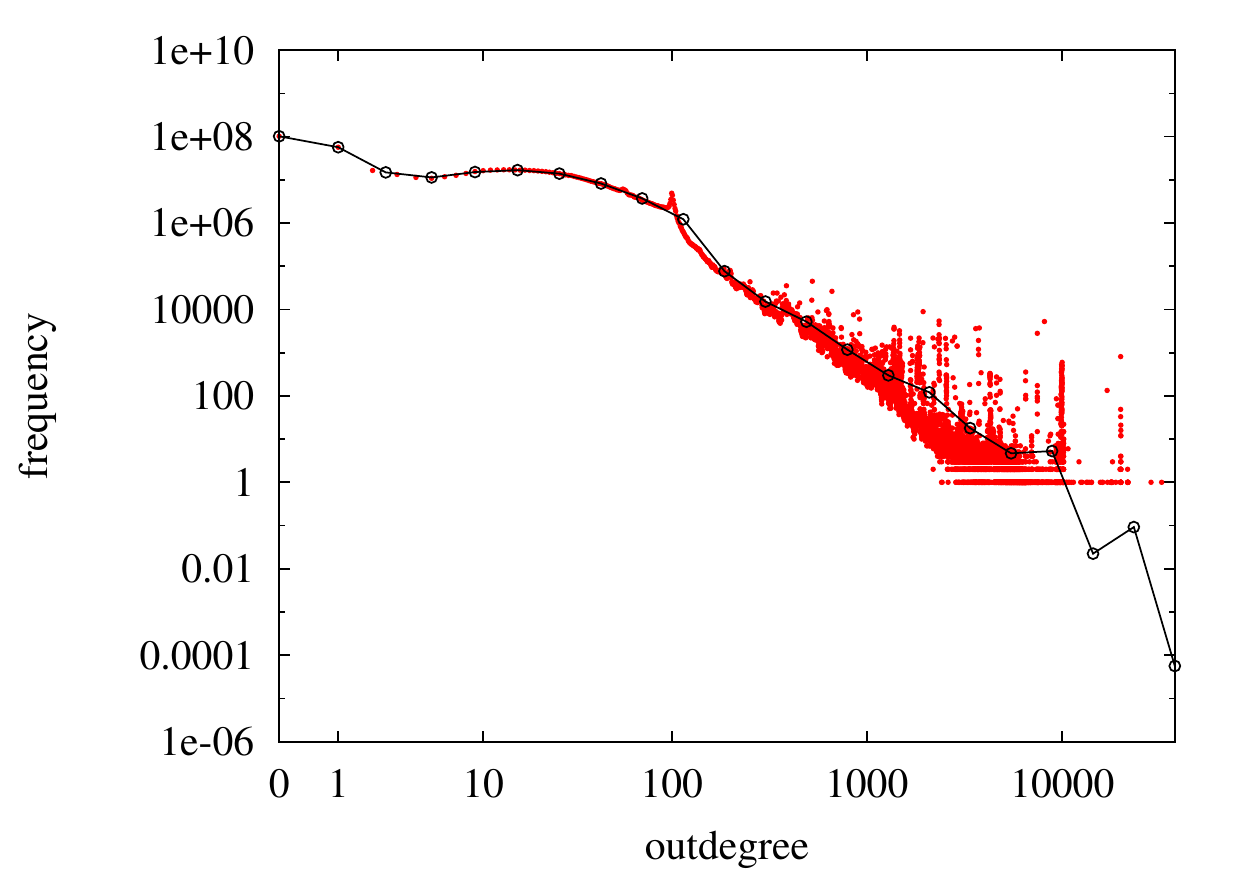}
\end{tabular}}
\caption{\label{fig:outdeg}Outdegree plots for \texttt{uk-2014}, \texttt{eu-2015} and \texttt{gsh-2015} (degree/frequency plots with Fibonacci binning).}
\end{figure}

\begin{figure}[tp]
\centering
\begin{tabular}{ccccc}
\includegraphics[scale=0.35]{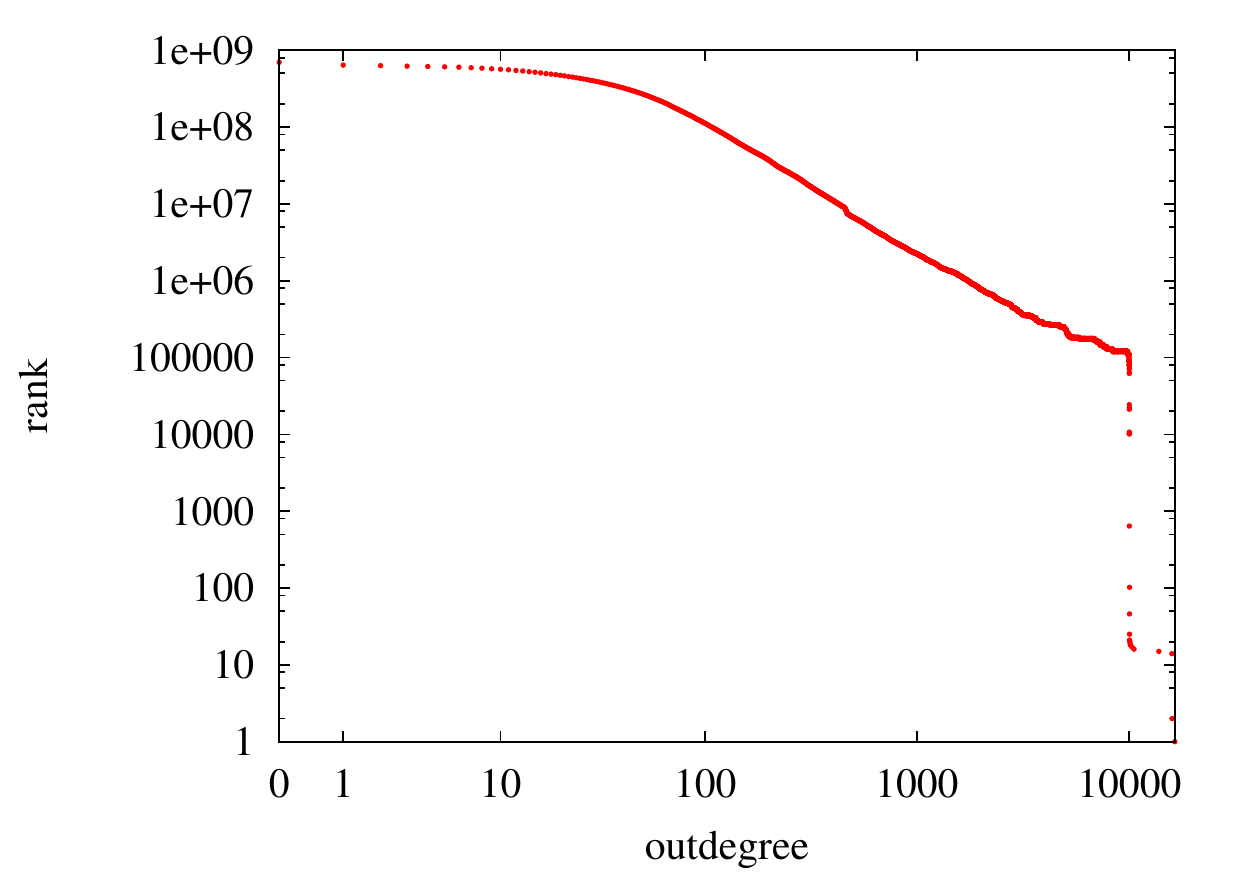} 
& &
\includegraphics[scale=0.35]{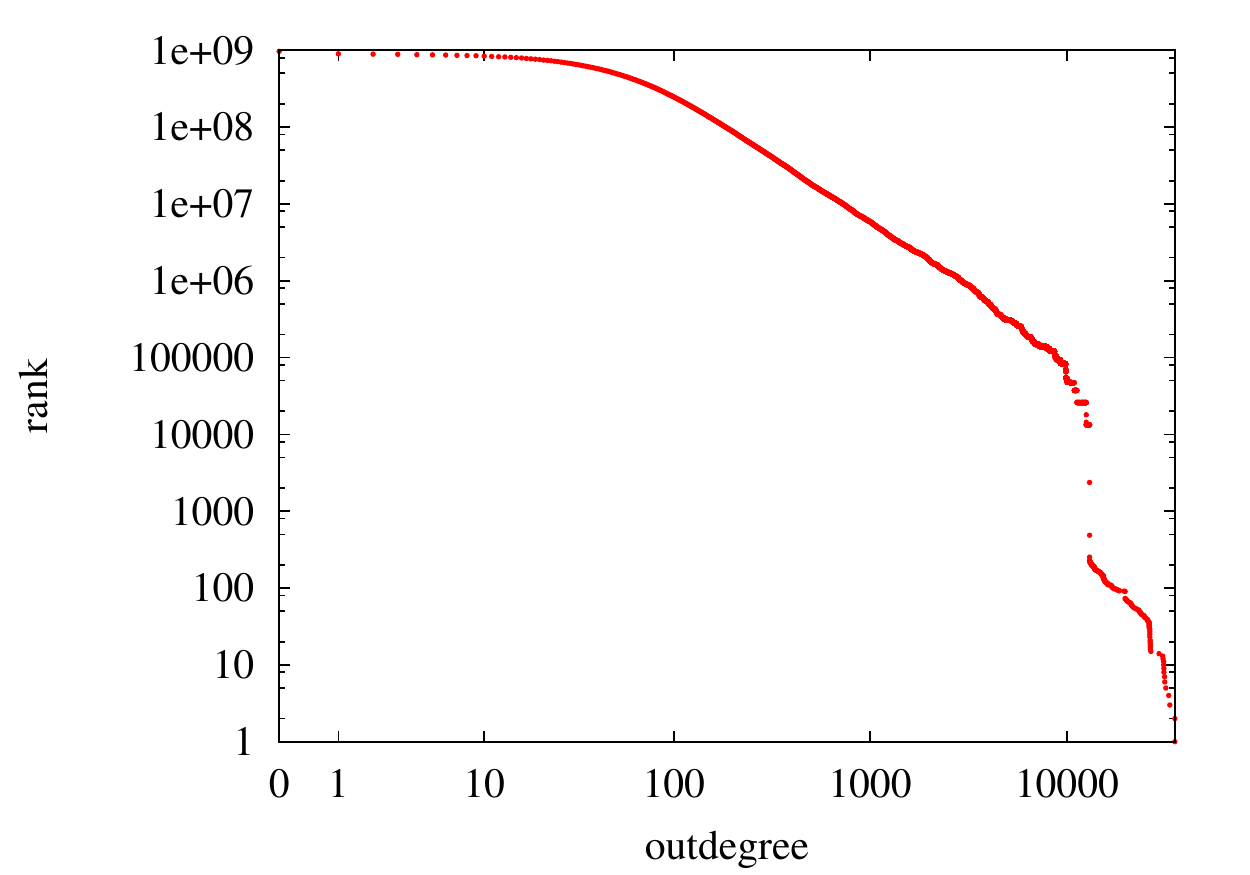}
& &
\includegraphics[scale=0.35]{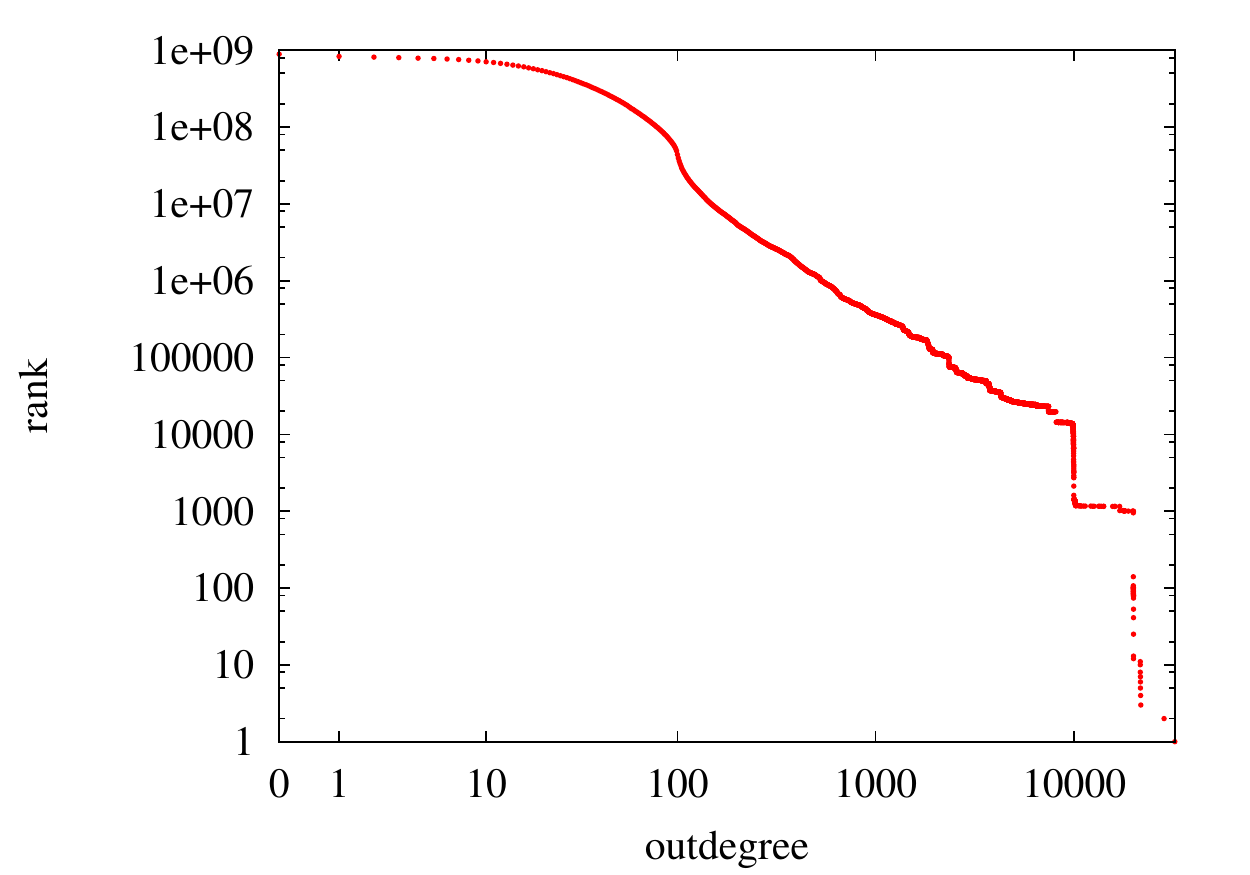}
\end{tabular}
\caption{\label{fig:outdeg-sr}Outdegree plots for \texttt{uk-2014}, \texttt{eu-2015} and \texttt{gsh-2015} (cumulative degree/rank plots).}
\end{figure}

The indegree and outdegree distributions are shown in Figures~\ref{fig:indeg},
\ref{fig:indeg-sr}, \ref{fig:outdeg} and \ref{fig:outdeg-sr}. We provide both a degree-frequency plot
decorated with \emph{Fibonacci binning}~\cite{VigFB}, and a degree-rank
plot\footnote{Degree-rank plots are the numerosity-based discrete analogous of
the complementary cumulative distribution function of degrees.
They give a much clearer picture than frequency dot plots when the data points
are scattered and highly variable.} to highlight with more precision the tail behaviour.

From Table~~\ref{tbl:basic}, we can see that pages at low depth tend to have
less outlinks, but more external links than inner pages. The content is
similarly smaller (content lives deeper in the structure of websites). Not
surprisingly, moreover, pages of the shallow snapshot are closer to one
another.


The most striking feature of the indegree distribution is an answer to
our question: \emph{the tail of the indegree distribution is, by and large,
shaped by the number of intra-host inlinks of root pages}. This is very visible
in the \texttt{uk-2014} snapshot, where limiting the host size at 10\,000
causes a sharp step in the degree-rank plot; and the same happens at 100 for
\texttt{gsh-2015}. But what is maybe even more interesting is that the visible
curvature of \texttt{eu-2015} is almost absent from \texttt{gsh-2015}. Thus, if
the latter (being mainly shaped by inter-host links) has some chance of being a
power-law, as proposed by the class of ``richer get richer'' models, the former
has none.
Its curvature clearly shows that the indegree distribution is not a power-law (a phenomenon already
noted in the analysis of the Common Crawl 2012 dataset~\cite{MVLGSW}): fitting it with the 
method by Clauset, Shalizi and Newman~\cite{CSNPLED} gives a p-value $<10^{-5}$  (and the same happens for the top-level domain graph).

\subsection{Centrality}

Table~\ref{tab:ukcen},~\ref{tab:eucen} and~\ref{tab:gshcen} report centrality
data about our three snapshots. Since the page-level graph gives rise to extremely
noisy results, we computed the \emph{host graph} and the
\emph{top-level domain} graph. In the first graph, a node is a host, and there
is an arc from host $x$ to host $y$ if some page of $x$ points to some page of
$y$. The second graph is built similarly, but now a node is a \emph{set of hosts} sharing the same
\emph{top-level domain} (TLD).
The TLD of a URL is determined from
its host using the \emph{Public Suffix List} published by the Mozilla
Foundation,\footnote{\url{http://publicsuffix.org/list/}} and it is
defined as one dot level above that the public suffix of the host: for
example, {\small\url{a.com}} for {\small\url{b.a.com}} (as
{\small\url{.com}} is on the public suffix list) and
{\small\url{c.co.uk}} for {\small\url{a.b.c.co.uk}} (as
{\small\url{.co.uk}} is on the public suffix list).\footnote{Top-level
domains have been called \emph{pay-level domain} in~\cite{MVLGSW}}

For each graph, we display the top ten nodes by indegree, PageRank (with
constant preference vector and $\alpha=0.85$) and by \emph{harmonic
centrality}~\cite{BoVAC}, the harmonic mean of all distance towards a node.
PageRank was computed with the highest possible precision in IEEE format using
the LAW library, whereas harmonic centrality was approximated using
HyperBall~\cite{BoVHB}.

Besides the obvious shift of importance (UK government sites for
\texttt{uk-2014}, government/news sites in \texttt{eu-2015} and large US
companies in \texttt{gsh-2015}), we con confirm the results of~\cite{MVLGSW}:
on these kinds of graphs, harmonic centrality is much
more precise and less prone to spam than indegree or PageRank. In the host
graphs, almost all results of indegree and most results of PageRank are spam or
service sites, whereas harmonic centrality identifies sites of interest (in
particular in \texttt{uk-2014} and \texttt{eu-2015}). At the TLD level, noise
decreases significantly, but the difference in behavior is still striking, with
PageRank and indegree still displaying several service sites, hosting providers
and domain sellers as top results.

\begin{table}
\renewcommand{\arraystretch}{1.2}
\tbl{\label{tbl:basic}Basic data}{
\centering
\begin{tabular}{l|r|r|r}
&\multicolumn{1}{c|}{\texttt{uk-2014}} & \multicolumn{1}{c|}{\texttt{gsh-2015}} & \multicolumn{1}{c}{\texttt{eu-2015}}\\
\hline
Overall	& 1\,477\,881\,641	&	1\,265\,847\,463	&	1\,301\,211\,841 \\
Archetypes & 787\,830\,045 & 1\,001\,310\,571 & 1\,070\,557\,254 \\
Avg. content length & 56\,039 &32\,526 &57\,027\\
Avg. outdegree & 105.86 & 96.34 & 142.60\\
Avg. external outdegree& 25.53 & 33.68 & 25.34\\
Avg. distance & 20.61 & 12.32& 12.45\\
Harmonic diameter & 24.63 & 14.91 & 14.18\\
Reachable pairs & 67.27\% & 80.29\% & 85.14\% \\
\end{tabular}}
\end{table}

\begin{sidewaystable}
\tbl{\label{tab:ukcen}Most relevant hosts and TPDs of \texttt{uk-2014} by different centrality measures.}{%
\renewcommand{\arraystretch}{1.3}
\centering
\begin{tabular}{|lr|lr|lr|}
\hline
\multicolumn{2}{|c|}{Indegree} & \multicolumn{2}{c|}{PageRank} &
\multicolumn{2}{c|}{Harmonic centrality}\\
\hline
\multicolumn{6}{|c|}{Host Graph}\\
\hline
postcodeof.co.uk &       726240 &            postcodeof.co.uk &       0.02988 &          postcodeof.co.uk &       1318833.61\\
namefun.co.uk &  661692 &                    namefun.co.uk &  0.02053 &                  www.google.co.uk &       1162789.19\\
www.slovakiatrade.co.uk &128291 &            www.slovakiatrade.co.uk &0.00543 &          www.nisra.gov.uk &       1130915.01\\
catalog.slovakiatrade.co.uk &    103991 &    catalog.slovakiatrade.co.uk &    0.00462 &  www.ons.gov.uk & 1072752.04\\
www.quiltersguild.org.uk &       93573 &     london.postcodeof.co.uk &0.00462 &          www.bbc.co.uk &  1067880.20\\
quiltersguild.org.uk &   93476 &             www.spanishtrade.co.uk & 0.00400 &          namefun.co.uk &  1057516.35\\
www.spanishtrade.co.uk & 87591 &             catalog.spanishtrade.co.uk &     0.00376 &  www.ordnancesurvey.co.uk &       1043468.95\\
catalog.spanishtrade.co.uk &     87562 &     www.germanytrade.co.uk & 0.00346 &          www.gro-scotland.gov.uk &1025953.27\\
www.germanytrade.co.uk & 73852 &             www.italiantrade.co.uk & 0.00323 &          www.ico.gov.uk & 1004464.11\\
catalog.germanytrade.co.uk &     73850 &     catalog.germanytrade.co.uk &    0.00322 &  www.nhs.uk &     1003575.06\\
\hline                  
\multicolumn{6}{|c|}{TPD Graph}\\
\hline
bbc.co.uk	&	60829	&	google.co.uk	&	0.00301	&	bbc.co.uk	&	422486.00\\
google.co.uk	&	54262	&	123-reg-expired.co.uk	&	0.00167	&	google.co.uk	&	419942.55\\
www.nhs.uk	&	22683	&	bbc.co.uk	&	0.00150	&	direct.gov.uk	&	375068.62\\
direct.gov.uk	&	20579	&	ico.gov.uk	&	0.00138	&	parliament.uk	&	371941.43\\
nationaltrust.org.uk	&	20523	&	freeparking.co.uk	&	0.00093	&	www.nhs.uk	&	370448.34\\
hse.gov.uk	&	13083	&	ico.org.uk	&	0.00088	&	ico.gov.uk	&	368878.14\\
timesonline.co.uk	&	11987	&	website-law.co.uk	&	0.00087	&	nationaltrust.org.uk	&	367367.47\\
amazon.co.uk	&	11900	&	hibu.co.uk	&	0.00085	&	telegraph.co.uk	&	364763.80\\
parliament.uk	&	11622	&	1and1.co.uk	&	0.00073	&	hmrc.gov.uk	&	364530.15\\
telegraph.co.uk	&	11467	&	tripadvisor.co.uk	&	0.00062	&	hse.gov.uk	&	361314.39\\
\hline                  
\end{tabular}}
\end{sidewaystable}

\begin{sidewaystable}
\tbl{\label{tab:eucen}Most relevant hosts and TPDs of \texttt{eu-2015} by different centrality measures.}{%
\renewcommand{\arraystretch}{1.3}
\centering
\begin{tabular}{|lr|lr|lr|}
\hline
\multicolumn{2}{|c|}{Indegree} & \multicolumn{2}{c|}{PageRank} &
\multicolumn{2}{c|}{Harmonic centrality}\\
\hline
\multicolumn{6}{|c|}{Host Graph}\\
\hline
www.toplist.cz  &       174433  &       www.myblog.de   &       0.001227        &       youtu.be        &       2368004.25\\
www.radio.de    &       139290  &       www.domainname.de       &       0.001215        &       ec.europa.eu    &       2280836.77\\
www.radio.fr    &       138877  &       www.toplist.cz  &       0.001135        &       europa.eu       &       2170916.37\\
www.radio.at    &       138871  &       www.estranky.cz &       0.000874        &       www.bbc.co.uk   &       2098542.10\\
www.radio.it    &       138847  &       www.beepworld.de        &       0.000821        &       www.spiegel.de  &       2082363.21\\
www.radio.pt    &       138845  &       www.active24.cz &       0.000666        &       www.google.de   &       2061916.72\\
www.radio.pl    &       138843  &       www.lovdata.no  &       0.000519        &       www.europarl.europa.eu  &       2050110.04\\
www.radio.se    &       138840  &       www.mplay.nl    &       0.000490        &       news.bbc.co.uk  &       2046325.37\\
www.radio.es    &       138839  &       zl.lv   &       0.000479        &       curia.europa.eu &       2038532.77\\
www.radio.dk    &       138838  &       www.mapy.cz     &       0.000472        &       eur-lex.europa.eu       &       2011251.37\\
\hline                  
\multicolumn{6}{|c|}{TPD Graph}\\
\hline
europa.eu       & 74129   &       domainname.de   & 0.001751   &       europa.eu       & 1325894.51\\
e-recht24.de    & 59175   &       toplist.cz      & 0.000700    &       youtu.be        & 1307427.57\\
youtu.be        & 47747   &       e-recht24.de    & 0.000688    &       google.de       & 1196817.20\\
toplist.cz      & 46797   &       mapy.cz & 0.000663    &       bbc.co.uk       & 1194338.96\\
google.de       & 40041   &       youronlinechoices.eu    & 0.000656   &       spiegel.de      & 1174629.32\\
mapy.cz & 38310   &       europa.eu       & 0.000640    &       free.fr & 1164237.86\\
google.it       & 35504   &       google.it       & 0.000444    &       bund.de & 1158448.65\\
phoca.cz        & 30339   &       youtu.be        & 0.000437   &       mpg.de  & 1155542.20\\
webnode.cz      & 28506   &       google.de       & 0.000420   &       admin.ch        & 1153424.50\\
free.fr & 27420   &       ideal.nl        & 0.000386  &       ox.ac.uk        & 1135822.35\\
\hline                  
\end{tabular}}
\end{sidewaystable}

\begin{sidewaystable}
\tbl{\label{tab:gshcen}Most relevant hosts and TPDs of \texttt{gsh-2015} by different centrality measures.}{%
\renewcommand{\arraystretch}{1.3}
\centering
\begin{tabular}{|lr|lr|lr|}
\hline
\multicolumn{2}{|c|}{Indegree} & \multicolumn{2}{c|}{PageRank} &
\multicolumn{2}{c|}{Harmonic centrality}\\
\hline
\multicolumn{6}{|c|}{Host Graph}\\
\hline
gmpg.org & 2423978      &       wordpress.org & 0.00885 &       www.google.com & 18398649.60 \\
www.google.com & 1787380        &       www.google.com & 0.00535        &       gmpg.org & 17167143.30 \\
fonts.googleapis.com & 1715958  &       fonts.googleapis.com & 0.00359  &       fonts.googleapis.com & 17043381.45 \\
wordpress.org & 1389348 &       gmpg.org & 0.00325      &       wordpress.org & 16326086.35 \\
maps.google.com & 959919        &       go.microsoft.com & 0.00317      &       play.google.com & 16317377.30 \\
www.miibeian.gov.cn & 955938    &       sedo.com & 0.00192      &       plus.google.com & 16300882.95 \\
www.adobe.com & 670180  &       developers.google.com & 0.00167 &       maps.google.com & 16105556.40 \\
go.microsoft.com & 642896       &       maps.google.com & 0.00163       &       www.adobe.com & 16053489.60 \\
www.googletagmanager.com & 499395       &       support.microsoft.com & 0.00146 &       support.google.com & 15443219.60 \\
www.blogger.com & 464911        &       www.adobe.com & 0.00138 &       instagram.com & 15262622.80 \\
\hline                  
\multicolumn{6}{|c|}{TPD Graph}\\
\hline
google.com & 2174980    &       google.com & 0.01011&   google.com & 10135724.15 \\
gmpg.org & 2072302      &       fonts.googleapis.com & 0.00628& gmpg.org & 9271735.90 \\
wordpress.org & 1409846 &       gmpg.org & 0.00611&     wordpress.org & 8936105.80 \\
fonts.googleapis.com & 1066178  &       sedo.com & 0.00369&     fonts.googleapis.com & 8689428.35 \\
adobe.com & 770597      &       adobe.com & 0.00307&    adobe.com & 8611284.30 \\
microsoft.com & 594962  &       wordpress.org & 0.00301&        microsoft.com & 8491543.60 \\
blogger.com & 448131    &       microsoft.com & 0.00277&        wordpress.com & 8248496.12 \\
wordpress.com & 430419  &       blogger.com & 0.00121&  yahoo.com & 8176168.72 \\
yahoo.com & 315723      &       networkadvertising.org & 0.00120&       creativecommons.org & 7985426.37 \\
statcounter.com & 313978        & 61.237.254.50 & 0.00105&    mozilla.org & 7960620.27 \\
\hline                  
\end{tabular}}
\end{sidewaystable}

\section{Conclusions}

In this paper we have presented \bubing, a new distributed
open-source Java crawler. \bubing is orders of magnitudes faster than existing
open-source crawlers, scales linearly with the number of agents, and will
provide the scientific community with a reliable tool to gather large data sets.

The main novel ideas in the design of \bubing are:
\begin{itemize}
  \item a pervasive usage of modern lock-free data structures to avoid contention among I/O-bound fetching threads;
  \item a new data structure, the \emph{workbench}, that is able to provide in constant
  time the next URL to be fetched respecting politeness both at the host and IP level;
  \item a simple but effective \emph{virtualizer}---a memory-mapped, on-disk store of 
  FIFO queues of URLs that do not fit into memory.
\end{itemize}


\bubing pushes software components to their limits by using massive parallelism
(typically, several thousand fetching threads); the result is a
beneficial fallout on all related projects, as witnessed by several
enhancements and bug reports to important software libraries like the Jericho HTML
parser and the Apache Software Foundation HTTP client, in particular in the area
of object creation and lock contention. In some cases, like a recent regression
bug in the ASF client (JIRA issue 1461), it was exactly \bubing's high
parallelism that made it possible to diagnose the regression.

Future work on \bubing includes integration with spam-detection software, and
proper handling of spider traps (especially, but not only, those consisting in
infinite non-cyclic HTTP-redirects); we also plan to implement policies for
IP/host politeness throttling based on download times and site branching speed,
and to integrate \bubing with different stores like HBase, HyperTable and
similar distributed storage systems.
As briefly mentioned, it is easy to let \bubing follow a different priority
order than breadth first, provided that the priority is \textit{per host} and
\textit{per agent}; the latter restriction can be removed at a moderate
inter-agent communication cost.
Prioritization at the level of URLs requires deeper changes in the inner
structure of visit states and may be implemented using, for example, the
Berkeley DB as a virtualizer: this idea will be a subject of future
investigations.

Another interesting direction is the integration with recently developed libraries
which provides \emph{fibers}, a user-space, lightweight alternative to threads that might
further increase the amount of parallelism available using our synchronous I/O design. 




\section{Acknowledgments}
 We thank our university for
providing bandwidth for our experiments (and being patient with bugged
releases). We thank Giuseppe Attardi, Antonio Cisternino and Maurizio Davini for
providing the hardware, and the GARR Consortium for providing the bandwidth for
experiments performed at the Universit\`a di Pisa. Finally, we thank Domenico
Dato and Renato Soru for providing the hardware and bandwidth for the iStella
experiments.

\bibliography{law}

\end{document}